# A comprehensive numerical model of wind-blown sand

**Jasper F. Kok[1,2,*], and Nilton O. Renno[1,2]**
[1]Applied Physics Program, University of Michigan, Ann Arbor, MI, 48109, USA.
[2]Atmospheric, Oceanic, and Space Sciences, University of Michigan, Ann Arbor, MI, 48109, USA.

# ABSTRACT

Wind-blown sand, or 'saltation', ejects dust aerosols into the atmosphere, creates sand dunes, and erodes geological features. We present a comprehensive numerical model of steady-state saltation that, in contrast to most previous studies, can simulate saltation over mixed soils. Our model simulates the motion of saltating particles due to gravity, fluid drag, particle spin, fluid shear, and turbulence. Moreover, the model explicitly accounts for the retardation of the wind due to drag from saltating particles. We also developed a physically-based parameterization of the ejection of surface particles by impacting saltating particles which matches experimental results. Our numerical model is the first to reproduce measurements of the wind shear velocity at the impact threshold (i.e., the lowest shear velocity for which saltation is possible) and of the aerodynamic roughness length in saltation. It also correctly predicts a wide range of other saltation processes, including profiles of the wind speed and particle mass flux, the total height-integrated mass flux, and the size distribution of saltating particles. Indeed, our model is the first to reproduce such a wide range of experimental data. Since we use a minimum number of empirical relations, our model can be easily adapted to study saltation under a variety of physical conditions, such as saltation on other planets, saltation under water, and saltating snow. We aim to use our model to develop a physically-based parameterization of dust emission for use in climate models.

# ARTICLE

## 1. Introduction

A comprehensive understanding of wind-blown sediment transport is essential for a wide range of problems across scientific disciplines. Indeed, wind-blown sand and dust creates sand dunes and dune ripples [*Bagnold*, 1941], erodes geological features [*Greeley and Iversen*, 1985] and agricultural fields [*Sterk*, 2003], and are a key component of dust storms and dust devils on Earth and Mars [*Shao*, 2000; *Renno et al.*, 2004]. Moreover, soil dust emitted into the atmosphere plays a major role in many earth system processes [*Goudie and Middleton*, 2006; *IPCC*, 2007], including by providing limiting micronutrients such as iron and phosphorus to a variety of ecosystems [*Jickells et al.*, 2005], serving as cloud nuclei [*DeMott et al.*, 2003], scattering and absorbing both shortwave and longwave radiation [*Sokolik et al.*, 2001], enhancing melting of snow packs and glaciers upon deposition [*Painter et al.*, 2007], and possibly affecting hurricane formation in the Atlantic Ocean [*Sun et al.*, 2008]. Mineral dust aerosols can be a significant hazard to human health [*Prospero*, 1999]. Finally, the transport of sediment by wind also shapes the landscape on Mars, Venus, and Titan [e.g., *Greeley and Iversen*, 1985], and dust aerosols are of major importance to the Martian climate [*Fenton et al.*, 2007].

As wind speed increases, sand particles of ~70-500 μm diameter are the first to be moved by wind and bounce along the surface in a series of hops [*Greeley and Iversen*, 1985; *Shao*, 2000]. This process is known as 'saltation' (Figure 1), and the impact of saltating particles on the soil bed can mobilize particles of a wide range of sizes. However, the acceleration of particles larger than ~500 μm is limited by their large inertia and these particles generally do not enter saltation [*Shao*, 2000]. Instead, they roll or slide along the surface, driven by impacts of saltating particles and direct wind pressure in a mode of transport known as 'creep' [*Bagnold*, 1941]. Dust particles, which are defined as particles with diameter < 62.5 μm [*Greeley and Iversen*, 1985], are not normally lifted by wind because of their substantial cohesive forces [*Shao and Lu*, 2000; *Kok and Renno*, 2006]. Instead, they are mobilized from the soil through impacts of saltating particles [*Shao et al*, 1993], after which they can be transported upwards by turbulent eddies. Dust particles smaller than ~20 μm can travel thousands of kilometers from their sources [*Gillette and Walker*, 1977], and affect the Earth system in a variety of manners as outlined above and in *Goudie and Middleton* [2006].

The transport of soil particles by wind can thus be separated into several physical regimes: long-term suspension (< 20 μm diameter), short-term suspension (~20 – 70 μm), saltation (~70 – 500 μm), and creep (> ~500 μm) [*Shao*, 2000]. Saltation is arguably the most important physical regime, because it occurs at the lowest wind speeds and causes the other modes of wind-blown sediment transport [*Shao*, 2000]. Saltation is initiated when the shear stress τ exerted by wind on the soil surface exceeds the threshold $\tau_t$ (~ 0.05 N/m$^2$ for loose sand) at which surface particles are lifted. *Bagnold* [1941] derived a simple empirical expression for the threshold wind shear velocity (also called the friction velocity) at which loose sand particles start to saltate,

$$u*_t \equiv \sqrt{\tau_t / \rho_a} = A\sqrt{\frac{(\rho_p - \rho_a)}{\rho_a} g D_p}, \tag{1}$$

where $A \approx 0.10$ is a dimensionless scaling parameter, $\rho_p$ and $\rho_a$ are the particle and fluid densities, $g$ is the gravitational constant, and $D_p$ is the diameter of a sphere with the same volume as the irregularly shaped sand particle. After saltation starts, the particles lifted from the surface exchange momentum with the wind. Upon striking the surface at angles of ~5-15° with the horizontal, these particles rebound at much steeper angles and thus larger vertical speeds [*Rice et al.*, 1995, 1996; *Wang*, 2008]. After a few hops, saltating particles can be sufficiently accelerated by wind drag to eject (or 'splash') other particles from the soil [*Bagnold*, 1973; *Ungar and Haff*, 1987]. Newly ejected particles are then accelerated by wind and eject more particles when impacting the surface. This causes an exponential increase in the number of saltating particles in the initial stages of saltation [*Anderson and Haff*, 1988, 1991; *Shao and Raupach*, 1992; *McEwan and Willetts*, 1993]. Indeed, the rapid exponential increase causes the flux of saltating particles to 'overshoot' the eventual steady-state mass flux [*Anderson and Haff*, 1988, 1991; *Shao and Raupach*, 1992; *McEwan and Willetts*, 1993], after which the momentum fluxes of the fluid and saltating particles reach an equilibrium. This equilibrium is determined by the finite amount of momentum available to be transferred from the wind to the saltating particles, such that the wind profile in the presence of saltation is often substantially reduced from that without saltation [e.g., *Owen*, 1964].

In steady-state saltation, surface particles are rarely lifted directly by fluid forces because the wind shear velocity at the surface is lower than the 'fluid threshold' given by Eq. (1). This sub-threshold wind shear at the surface occurs because the transfer of momentum to the soil surface is dominated by the impacts of saltating particles, not by wind drag [*Bagnold*, 1937, 1973; *Ungar and Haff*, 1987; *Anderson and Haff*, 1988, 1991; *Shao and Raupach*, 1992; *McEwan and Willetts*, 1991, 1993]. As a result, once saltation is initiated, it can be maintained at shear velocities somewhat below the fluid threshold. The minimum shear velocity at which saltation can occur in this manner is termed the 'impact threshold' [*Bagnold*, 1941] and, for Earth ambient conditions, is approximately 80-85 % of the 'fluid threshold' defined by Eq. (1) [*Bagnold*, 1937].

Numerical models of the different physical processes involved in saltation have been developed by various researchers. *White and Schulz* [1977], *Hunt and Nalpanis* [1985], and *Anderson and Hallet* [1986] were the first to successfully model the trajectories of saltating particles. Building on the success of these initial studies, *Ungar and Haff* [1987] were the first to couple the motion of saltating particles to the retardation of the wind speed near the surface in a simple, steady-state model, in which the trajectories of all saltating particles were assumed identical. Nonetheless, they were able to reproduce some essential features of saltation, such as the near-surface focusing of the wind profiles for different shear velocities (first reported by *Bagnold* [1936]). *Werner* [1990] developed a more comprehensive numerical model of steady-state saltation that allowed for a range of particle trajectories. This model also included a parameterization of the splashing of particles from the soil, which was based on laboratory measurements of particle ejections [*Werner*, 1987]. However, only the more detailed models developed by *Anderson and Haff* [1988, 1991] and *McEwan and Willetts* [1991, 1993] were able to simulate the development of saltation from inception to steady-state. *Shao and Li* [1999] built on these studies and developed a saltation model as part of a large eddy model that explicitly solved for the wind field. More recently, *Almeida et al.* [2006] coupled a saltation model to the computational fluid dynamics model FLUENT capable of calculating the turbulent wind field in the presence of saltation. While their saltation model assumes identical trajectories and does not explicitly consider the 'splashing' of surface particles, they were able to reproduce empirical expressions for the saltation mass flux. They used this model to study saltation on Mars [*Almeida et al.*, 2008]. Zheng and co-workers also developed a numerical model that can reproduce certain essential features of saltation, and were the first to account for the effects of electrostatic forces [*Zheng et al.*, 2006]. The subsequent study of *Kok and Renno* [2008] indicated that electrostatic forces increase the saltating particle concentration [*Kok and Renno*, 2006] and lower the height of saltation trajectories, thereby possibly resolving the discrepancy between the measured [*Greeley et al.*, 1996; *Namikas*, 2003] and predicted [*Bagnold*, 1941; *Owen*, 1964] height of the saltation layer.

While the models discussed above have provided significant advancements in our understanding of the physics of saltation, they have been unable to accurately reproduce natural saltation. Moreover, these models were generally constrained to saltation of monodisperse particles, while natural saltation occurs over soils containing a range of particle sizes. We here present a comprehensive physically-based numerical model of saltation that can simulate saltation of soils consisting of particles of various sizes.

Our model shows reasonable to excellent agreement with a variety of experimental measurements such as the impact threshold, horizontal and vertical profiles of particle mass flux, the wind profile in saltation, and the size distribution of saltating particles. Our model also includes a physically-based parameterization of the splashing of surface particles, which agrees with available laboratory and numerical experiments. To the best of our knowledge, our numerical model is the first capable of reproducing such a wide range of experimental data. Since we use a minimum of empirical relations, we argue that our model can accurately simulate saltation in a variety of physical environments, including other planets such as Mars [*Kok and Renno*, 2008b; *Renno and Kok*, 2008; *Almeida et al.*, 2008], saltating snow, and saltation in water. Our model was coded in MATLAB and is freely available by contacting the first author (J.K.).

We describe our model in detail in the next section, compare its results to measurements in Section 3, and present conclusions in Section 4.

**2. Model description**

We model saltation as the interplay of several processes [*Werner*, 1990; *Anderson and Haff*, 1991; *McEwan and Willetts*, 1991]: (i) the motion of saltating particles, (ii) the modification of the wind profile through momentum transfer between the wind flow and saltating particles, and (iii) the collision of particles with the soil surface and the subsequent ejection or 'splashing' of surface particles into the fluid stream. For simplicity, we simulate particle motion in two dimensions, as previous investigators have also done [e.g., *Werner*, 1990; *Anderson and Haff*, 1991; *McEwan and Willetts*, 1991]. We also neglect the collisions of saltating particles with each other, as well as the effects of electrostatic forces on particle trajectories [*Kok and Renno*, 2006, 2008]. The effect of both these processes is limited for small to medium shear velocities (i.e., $u* < \sim 0.5$ m/s) but probably becomes important for larger shear velocities [*Kok and Renno*, 2008; *Sorensen and McEwan*, 1996; *Huang et al.*, 2007]. We therefore plan to include both mid-air collisions and electrostatic forces in a future model version [*Kok and Renno*, 2009].

Our numerical model simulates saltation in steady-state (see Figure 2). The model uses the logarithmic wind profile known as the "law of the wall" [Prandtl, 1935] to calculate the initial trajectories of saltating particles. The drag exerted by the particles on the wind is then obtained from these trajectories and used to adjust the wind profile. The concentration of saltating particles is calculated using the steady-state condition that the number of particles striking the soil must be equal to the number of rebounding and ejected particles. If the number of rebounding and ejected particles is greater than the number of impacting particles, then the concentration of saltating particles is increased accordingly, which enhances the exchange of momentum with the wind and reduces the near-surface wind speed, causing particles to strike the soil at reduced speed and thereby eject fewer particles. Due to their interdependence, the particle concentration, wind profile, and particle trajectories are calculated iteratively until steady-state is reached (see Figure 2). Because the interaction of saltating particles with the soil surface and the turbulent wind is stochastic (see Sections 2.1.2 and 2.2), these processes cause variability in the model simulations that can be seen as characteristic of natural saltation. 'Steady-state' saltation as simulated by our model thus entails a dynamic balance that, averaged over many iterations, satisfies the condition that the number of impacting particles is equal to the number of particles that rebound and are ejected from the soil [*Jackson and McCloskey*, 1997; *Anderson and Haff*, 1991].

We discuss each component of the model in detail below. Where possible, we use experimental results to verify the performance of individual model components.

**2.1 Particle trajectories**

The motion of saltating particles is determined mainly by gravitational and fluid forces. For the present model version, we thus neglect electrostatic forces [*Kok and Renno*, 2008] and mid-air collisions [*Sorensen and McEwan*, 1996; *Dong et al.*, 2005; *Huang et al.*, 2007] which affect particle trajectories mostly for large shear velocities (see Section 2.6).

**2.1.1 Fluid forces**

The main fluid force affecting particle trajectories is the drag force [e.g., *Anderson and Haff*, 1991],

$$F_d = -\frac{\pi D_p^2}{8} \rho_a C_d v_R \mathbf{v_R}, \tag{2}$$

where $\mathbf{v_R} = \mathbf{v} - \mathbf{U}$ is the difference between the particle (**v**) and wind (**U**) velocities, and $v_R = |\mathbf{v_R}|$. The drag coefficient ($C_d$) of natural sand particles is generally larger than that for spherical particles of the same volume, both because their irregular shape produces a larger effective surface area than a sphere and because regions of large curvature can lead to flow separation, which increases the drag [*Dietrich*, 1982]. Detailed measurements of the terminal velocity in water have been used to measure the drag coefficient of natural sand particles [*Dietrich*, 1982; *Camenen*, 2007]. We calculate the drag coefficient of a saltating sand particle using an equation proposed by *Cheng* [1997] that includes the effects discussed above

$$C_d = \left[\left(\frac{32}{\text{Re}}\right)^{2/3} + 1\right]^{3/2}, \quad (3)$$

where the particle Reynolds number is given by

$$\text{Re} = \frac{\rho_a v_R D_p}{\mu}. \quad (4)$$

Saltating particles also experience lift forces both due to the shearing flow (the 'Saffman force') [*Saffman*, 1965, 1968], and from particle rotation (the 'Magnus force') [*Rubinow and Keller*, 1961]. We calculate these lift forces using the following expressions proposed by *Loth* [2008]

$$F_{\text{saff}} = 1.615 J^* D_p^2 \left(\rho_a \mu \frac{\partial U_x}{\partial z}\right)^{1/2} (\hat{\mathbf{y}} \times \mathbf{v_R}), \text{ and} \quad (5)$$

$$F_{\text{mag}} = \frac{\pi}{8} \rho_a D_p^3 C_{L\Omega}^* (\mathbf{\Omega}_p \times \mathbf{v_R}), \quad (6)$$

where $U_x$ is the horizontal wind speed, $\hat{\mathbf{y}}$ is the unit vector perpendicular to the plane in which particle motion takes place, and $J^*$ is a strong function of the shear of the flow, the kinematic viscosity, and the relative velocity of the particle to the fluid, and is defined by *McLaughlin* [1991]. The normalized spin lift coefficient $C_{L\Omega}^*$ is given by Eq. 16 in *Loth* [2008] and is ~0.5-0.7 for normal flow conditions in saltation on Earth. Previous studies have assumed $C_{L\Omega}^* = 1$, which is a good approximation only for Re << 1 [*Rubinow and Keller*, 1961; *White and Schulz*, 1977, *Loth*, 2008], and thus overestimates the lift force caused by particle spin [*Hunt and Nalpanis*, 1985; *Shao*, 2000]. The particle angular velocity $\Omega_p$ is defined as positive for topspin (i.e., the particle rotates as if rolling in the same direction as it is moving), in which case the lift force is also positive (i.e., pointing upwards). Experiments have shown that saltating particles predominantly have topspin, with $\Omega_p$ in the range of 100 - 1000 rev/s [*Chepil and Woodruff*, 1963; *White and Schulz*, 1977; *White*, 1982; *Xie et al.*, 2007; *Zou et al.*, 2007]. A likely reason for the predominance of topspin is that the shearing flow exerts a moment on the particles that produces topspin. Moreover, the friction on a particle's underside upon collision with the soil surface also produces torques that favor topspin over backspin. We assume that, after colliding with the surface, saltating particles have an initial spin of $\Omega_{p,0} \approx 400 \pm 500$ rev/s, as suggested by experiments [*Chepil and Woodruff*, 1963; *White and Schulz*, 1977; *White*, 1982; *Xie et al.*, 2007; *Zou et al.*, 2007]. After leaving the surface, the particle spin is affected by the shear of the flow (which imparts topspin), and by viscous dissipation (which reduces the particle spin). Thus, after stochastically determining the particle's spin upon leaving the surface, we calculate the particle spin as a function of time by numerically integrating the differential equation [*Anderson and Hallet*, 1986; *Loth*, 2008]

$$\frac{d\Omega_p}{dt} = \frac{60\mu}{\rho_p D_p^2}\left(\frac{1}{2}\frac{\partial U_x}{\partial z} - \Omega_p\right), \tag{7}$$

where the first term in the brackets on the right-hand side represents the moment exerted by the shearing flow, and the second term denotes viscous dissipation. We neglect forces due to particle rotation that are not in the *xz*-plane (see Figure 1) [*Xie et al.*, 2007].

The Saffman force due to the shearing flow is very small, except very close to the surface where the shear is large. In fact, sensitivity studies with our model indicate that the Saffman force can be neglected without measurably affecting particle trajectories. The Magnus lift force due to particle rotation has typical values of a few percent of the particle's weight and therefore does significantly affect particle trajectories, as also indicated by laboratory studies [*White and Schulz*, 1977; *White*, 1982; *Zou et al.*, 2007].

**2.1.2 Effect of turbulence on particle trajectories**

Previous numerical models of saltation have often neglected the effects of turbulence on particle trajectories [e.g., *Anderson and Haff*, 1988; *McEwan and Willetts*, 1991], despite the fact that turbulence can significantly affect the trajectories of particles smaller than ~250 μm [*Anderson*, 1987]. We therefore do include the effects of turbulence on particles trajectories.

The wind speed can be decomposed into the average wind speed and the turbulent fluctuation:

$$U_x = \overline{U_x} + U_x'; \quad U_z = \overline{U_z} + U_z' \tag{8}$$

where $\overline{U_x}$, $\overline{U_z}$, $U_x'$, and $U_z'$ are respectively the time-averaged and turbulent horizontal and vertical components of the wind speed at a given height. In the case studies presented in this article, we assume horizontal flow (i.e., $\overline{U_z} = 0$). The calculation of $\overline{U_x}$ in the near-surface layer where saltation takes place (the 'saltation layer') is discussed in Section 2.3. The turbulent fluctuation experienced by a fluid parcel moving with the flow can be described statistically by [*Van Dop et al.*, 1985; *Wilson and Sawford*, 1996]

$$U_z'(t+dt) - U_z'(t) = -\frac{U_z'(t)}{T_L}dt + n_G \sigma_w \sqrt{2dt/T_L}, \tag{9}$$

where a similar equation describes $U_x'$. Equation (9) has the discretized solution

$$U_z'(t+\Delta t) = U_z'(t)\exp(-\Delta t/T_L) + n_G \sigma_w \sqrt{2}\left[1-\exp\left(-\sqrt{\Delta t/T_L}\right)\right], \tag{10}$$

which in the limit $\Delta t \to dt$ reduces to Eq. (9). The model time step $\Delta t$ is always set smaller than the Lagrangian time scale ($T_L$), and $n_G$ is a Gaussian distributed random number with zero mean and unit standard deviation. For homogeneous, isotropic turbulence, the standard deviations of the horizontal and vertical turbulent wind speeds equal

$$\sigma_u = b_u \kappa z \left(\frac{\partial \overline{U_x}}{\partial z}\right) \; ; \; \sigma_w = b_w \kappa z \left(\frac{\partial \overline{U_x}}{\partial z}\right), \tag{11}$$

where $b_u = 1.4 \pm 0.1$ and $b_w = 2.5 \pm 0.1$ [*Hunt and Weber*, 1979; *Shao*, 1995; *Nishimura and Hunt*, 2000], and where $\kappa = 0.40$ is the von Kármán constant. The Lagrangian time scale $T_L$ represents the approximate time scale over which the velocities experienced by a fluid parcel at times $t$ and $t + T_L$ are statistically related. Since measurements are generally made in a stationary frame of reference, it is notoriously difficult to measure the Lagrangian time scale [*Leuning*, 2000]. To the best of our knowledge, there have been no detailed studies of this time scale in saltation layers. However, the Lagrangian time scale of turbulent flow in forest and vegetation canopies has been studied in detail [*Raupach et al.*, 1996; *Leuning et al.*, 2000]. We thus use the analogy between turbulent flows in forest canopies and saltation layers [*Raupach*, 1991], and define $T_L$ following Eqs. (10) and (11) of *Leuning et al.* [2000] by equating the canopy height $h_c$ to the height below which the bulk (i.e., 95 %) of the saltation mass flux occurs.

Equations (9)-(11) describe the turbulent fluctuation of the wind speed experienced by a particle that exactly follows the wind flow. However, gravitational forces and inertia cause the movement of saltating particles to deviate from that of the fluid [*Anderson*, 1987; *Sawford and Guest*, 1991]. The time scale $T_L^*$ over which the fluctuations in wind speeds experienced by a saltating particle remain statistically correlated is thus shorter [*Csanady*, 1963], because a particle with non-zero velocity relative to the flow requires less time to traverse a turbulent eddy. Although these effects are still not fully understood [*Reynolds*, 2000], *Sawford and Guest* [1991] showed that a reasonable approximation for $T_L^*$ for use with the fluctuation of the vertical flow speed is

$$T_L^* = T_L \left[1 + (\beta v_R / \sigma_w)^2\right]^{-1/2}, \tag{12}$$

where $\beta = T_L/T_E$ is the ratio of the Lagrangian and Eulerian time scales, which is uncertain but is of order unity [*Sawford and Guest*, 1991; *Reynolds*, 2000; *Anfossi et al.*, 2006]. For horizontal velocity components (i.e., perpendicular to gravity),

$$T_L^* = T_L \left[1 + (2\beta v_R / \sigma_u)^2\right]^{-1/2}. \tag{13}$$

To test the accuracy of Eqs. (11) – (13), we used our model to simulate wind tunnel measurements of the dispersion of solid particles (see Figure 3) [*Snyder and Lumley*, 1971]. As in *Sawford and Guest* [1991], we found poor agreement between our model and the results of experiments for the lightest particle (47 μm hollow glass), but found excellent agreement for the heavier particles (47 μm copper, 87 μm glass, and 87 μm corn pollen). Since the weight and relaxation time of particles that show good agreement are similar to those of saltating particles, we use the above parameterization in our model.

We neglect the effect of saltating particles on the turbulence level (i.e., $\sigma_u$ and $\sigma_w$), because measurements indicate that such effects are small [*Taniere et al.*, 1997; *Nishimura and Hunt*, 2000].

### 2.1.3 Full equations of motion

We simulate the particle trajectories due to the gravitational and fluid forces described above. The full equations of motion are:

$$ma_x = -\frac{\pi}{8}D_p^2\rho_a\left[C_d v_R(v_x - U_x) + D_p C_{L\Omega}^*\Omega_p(v_z - U_z) + \frac{12.92}{\pi}J^*\left(\frac{\mu}{\rho_a}\frac{\partial \overline{U_x}}{\partial z}\right)^{1/2}(v_z - U_z)\right], \text{ and} \quad (14a)$$

$$ma_z = -\frac{\pi}{8}D_p^2\rho_a\left[C_d v_R(v_z - U_z) + D_p C_{L\Omega}^*\Omega_p(v_x - U_x) + \frac{12.92}{\pi}J^*\left(\frac{\mu}{\rho_a}\frac{\partial \overline{U_x}}{\partial z}\right)^{1/2}(v_x - U_x)\right] - mg, \quad (14b)$$

where $m$ is the particle's mass, $v_x$, $v_z$, $a_x$, and $a_z$ are respectively the particle speeds and accelerations in the $x$ and $z$ directions, and $g = 9.8$ m/s$^2$ is the gravitational constant. The first term on the right-hand side accounts for fluid drag, the second for particle spin, and the third for the Saffman force. The model uses the 4$^{th}$-order Adams-Moulton method [*Hairer et al.*, 1993] to numerically integrate the equations of motion and obtain the particle trajectories. In order to lower the computational cost, the model randomly selects a number of saltating particles specified by the user, calculates their trajectories, and considers those to represent the entire ensemble of saltating particle trajectories. Increasing the number of simulated saltating particle trajectories beyond the number used to obtain the results presented in this article does not significantly affect the model results.

## 2.2 Particle collisions with the surface

The collision of saltating particles with the surface (Figure 1) is a key physical process in saltation [*Anderson and Haff*, 1991; *Shao*, 2000]. Saltating particles strike the soil nearly horizontally, at ~5-15° from horizontal, and generally rebound at angles of ~15-70° from horizontal [*Anderson and Haff*, 1988, 1991; *McEwan and Willetts*, 1985, 1986, 1989; *Nalpanis et al.*, 1993; *Rice et al.*, 1995]. The collision of saltating particles with the soil thus converts horizontal momentum into vertical momentum [*Wang et al.*, 2008]. This is essential, as it allows saltating particles to replenish the vertical momentum that is dissipated through fluid drag. Moreover, particles striking the soil can dislodge and eject particles from the surface in a process termed 'splashing' [*Bagnold*, 1973; *Ungar and Haff*, 1987].

### 2.2.1 The rebounding particle

While particle trajectories can be calculated based on simple physical principles (see section 2.1), the collision of saltating particles with the soil surface is inherently a stochastic process. For example, not all saltating particles rebound from the surface, even when they impact it at high speed [*Mitha et al.*, 1986; *Anderson and Haff*, 1991]. The probability that a saltating particle will rebound upon impact can be approximated by [*Anderson and Haff*, 1991]

$$P_{\text{reb}} = B[1 - \exp(-\gamma v_{\text{imp}})], \tag{15}$$

where $v_{\text{imp}}$ is the speed with which the particle impacts the surface. *Mitha et al.* [1986] determined the parameter $B$ to be 0.94 for 4 mm steel particles, while the 2-dimensional numerical simulations of *Anderson and Haff* [1991] found a similar value of $B \approx 0.95$ for 230 and 320 μm sand particles. To the best of our knowledge, the parameter $\gamma$ has not been experimentally determined, but the numerical simulations of *Anderson and Haff* [1988, 1991] indicate that it is of order 2 s/m.

We use results of laboratory and numerical studies to describe the velocity of rebounding particles [*White and Schulz*, 1977; *Mitha at al.*, 1986; *Anderson and Haff*, 1991; *McEwan and Willetts*, 1991; *Nalpanis et al.*, 1993; *Rice et al.*, 1995; *Rioual et al.*, 2000; *Oger et al.*, 2005; *Beladjine et al.*, 2007; *Kang et al.*, 2008]. Recent laboratory experiments have shown that the fraction of kinetic energy retained by the rebounding particle is approximately normally distributed [*Wang et al.*, 2008] while the rebounding angle approximately follows an exponential distribution [*Kang et al.*, 2008; *Willetts and Rice*, 1985, 1986; *McEwan and Willetts*, 1991; *Rice et al.*, 1996]. We thus take the kinetic energy of the rebounding particles to be 45 ± 22 % of the impacting kinetic energy, and the rebound angle as an exponential distribution with a mean of 40º from horizontal.

**2.2.2 Ejection speed of splashed surface particles**

In steady-state saltation, the loss of particles through the process represented by Eq. (15) is balanced by the 'splashing' of surface particles. The 'splash function,' which describes the number and velocity of the ejected surface particles as a function of the velocity of the impacting particle [*Ungar and Haff*, 1987] is thus a key component of numerical models of saltation [*Werner*, 1990; *Anderson and Haff*, 1988, 1991; *McEwan and Willetts*, 1991, 1993; *Shao and Li*, 1999]. Instead of using an empirical expression for the splash function that is based on the results of laboratory or numerical experiments, as most previous models have done, we derive a physically based expression of the splash function below.

The ejection of particles from the surface by impacting saltating particles is constrained by the conservation of both energy and momentum. These constraints can be expressed as

$$\varepsilon_{\text{reb}} + \varepsilon_{\text{ej}} + \varepsilon_{\text{F}} = 1, \text{ and} \tag{16a}$$

$$\alpha_{\text{reb}} + \alpha_{\text{ej}} + \alpha_{\text{F}} = 1, \tag{16b}$$

where $\varepsilon$ and $\alpha$ respectively refer to the partitioning of energy and momentum, and the subscripts refer to the fraction of the total energy or momentum contained in the rebounding particle (reb), the ejected particles (ej), and that lost through frictional processes (F). In order to derive a physically-based expression of the number and speed of ejected particles, we need to determine whether energy conservation or momentum conservation is the dominant constraint on the ejection of surface particles. To determine this, we unrealistically neglect friction (i.e., $\varepsilon_F = \alpha_F = 0$) in the collision of a particle of mass $m_{imp}$ with a bed of particles with mass $m_{ej}$, such that we can obtain the maximum number of particles that can be ejected without violating conservation of energy ($N_{max}^E$) or momentum ($N_{max}^M$). This yields

$$N_{max}^E = \frac{(1-\varepsilon_{reb})m_{imp}v_{imp}^2}{m_{ej}\langle v_{ej}^2\rangle + 2\phi}, \text{ and} \tag{17a}$$

$$N_{max}^M = \frac{(1-\alpha_{reb})m_{imp}v_{imp}}{m_{ej}\langle v_{ej}\rangle}, \tag{17b}$$

where $\phi$ is the energy with which soil particles are bonded with each other, $\langle v_{ej}\rangle$ is the ensemble-averaged ejected particle speed (that is, the speed of ejected particles averaged over many impacts on the soil surface of a particle with a given speed), and $\langle v_{ej}^2\rangle$ is the ensemble-averaged square of the ejected particle speed.

In order to compare $N_{max}^E$ and $N_{max}^M$ we need to relate $\langle v_{ej}^2\rangle$ to $\langle v_{ej}\rangle$. Such a relation can be obtained by assuming a functional form for the probability distribution $P(v_{ej})$ of the speed of ejected particles. The numerical simulations of *Anderson and Haff* [1991] found that $P(v_{ej})$ takes the form of an exponential distribution, which is also suggested by experimental results (see Figure 4). We thus take [*Werner*, 1990; *Sorensen*, 1991; *Anderson and Haff*, 1991],

$$P(v_{ej}) = \frac{\exp(-v_{ej}/\langle v_{ej}\rangle)}{\langle v_{ej}\rangle}. \tag{18}$$

We find from Eq. (18) that $\langle v_{ej}^2\rangle = 2\langle v_{ej}\rangle^2$, which we combine with Eq. (17) to obtain the critical impact speed $v_{imp}^{crit}$ at which the constraints posed by energy and momentum conservation are equally restricting (i.e., where $N_{max}^E = N_{max}^M$). This yields

$$v_{imp}^{crit} = \frac{2}{1+\alpha_{reb}}\left[\langle v_{ej}\rangle + \phi/m_{ej}\langle v_{ej}\rangle\right] \approx \frac{2\langle v_{ej}\rangle}{1+\alpha_{reb}}, \tag{19}$$

where we used that $\varepsilon_{\text{reb}} = \alpha_{\text{reb}}^2$ and assumed that $\phi \ll m_{\text{ej}} \langle v_{\text{ej}} \rangle^2$ for loose sand, as is typical for saltation on dry dunes and beaches. When $v_{\text{imp}} \ll v_{\text{imp}}^{\text{crit}}$, we have that $N_{\text{max}}^E \ll N_{\text{max}}^M$, such that energy conservation constrains the number of surface particles that can be ejected. Conversely, when $v_{\text{imp}} \gg v_{\text{imp}}^{\text{crit}}$, we find that $N_{\text{max}}^E \gg N_{\text{max}}^M$, such that momentum conservation becomes the main constraint. Since the speed of ejected particles is approximately an order of magnitude smaller than the impacting speed [e.g., *Rice et al.*, 1995], we find that generally $v_{\text{imp}} \gg v_{\text{imp}}^{\text{crit}}$ and thus that $N_{\text{max}}^E \gg N_{\text{max}}^M$. This implies that the splashing of loose sand particles from the surface by saltating particles is limited primarily by momentum conservation, and not as much by energy conservation. While the inclusion of frictional processes will affect the exact value of $v_{\text{imp}}^{\text{crit}}$, it is unlikely to alter this general conclusion. Note however that the ejection of dust particles from the soil is rather different, because in this case $\phi$ is not small. Therefore, energy conservation might be the dominant constraint limiting the number of ejected dust particles. Indeed, this is what measurements by *Shao et al.* [1993] suggest.

We thus impose conservation of momentum on the number of surface particles that can be ejected, and thereby find that

$$N(v_{\text{imp}}) m_{\text{ej}} \langle v_{\text{ej}} \rangle = \langle \alpha_{\text{ej}} \rangle m_{\text{imp}} v_{\text{imp}}, \tag{20}$$

where $\langle \alpha_{\text{ej}} \rangle$ is the ensemble-averaged fraction of the impacting momentum that is spent on splashing particles from the surface, and $N$ is the average number of ejected particles, which depends on the particle impact speed $v_{\text{imp}}$. We neglect the dependence of $N$ on the impact angle [*Beladjine et al.*, 2007], because the range of angles with which saltating particles impact the surface is relatively narrow [e.g., *Wang et al.*, 2008]. Both laboratory and modeling studies suggest that the number of ejected particles scales approximately linearly with the impact speed [*Anderson and Haff*, 1988, 1991; *McEwan and Willetts*, 1991; *Rice et al.*, 1996; *Rioul et al.*, 2000; *Oger et al.*, 2005; *Beladjine et al.*, 2007],

$$N \approx A v_{\text{imp}}. \tag{21}$$

Dimensional analysis [*Beladjine et al.*, 2007] and conservation of momentum suggests that the parameter $A$ can be rewritten as

$$A = \frac{a}{\sqrt{gD}} \frac{m_{\text{imp}}}{m_{\text{ej}}}, \tag{22}$$

where $D$ is a typical particle size (~250 μm for saltation on Earth), and $a$ is a dimensionless constant that is independent of the impacting velocity and the masses of the impacting and ejected particles, and lies in the range of 0.01 – 0.05 [*Willetts and Rice*, 1985, 1986, 1989; *McEwan and Willetts*, 1991; *Rice et al.*, 1995, 1996]. Combining Eqs. (20)-(22) then yields the simple expression

$$\langle v_{ej}\rangle = \frac{\langle\alpha_{ej}\rangle\sqrt{gD}}{a}. \tag{23}$$

Thus, assuming that the fraction of momentum spent on splashing particles from the surface ($\langle\alpha_{ej}\rangle$) does not depend on impact speed, the average speed of ejected particles should be independent of the impact speed. This is indeed consistent with results for large impact speeds from laboratory experiments; Werner [1987, 1990] found that $\langle v_{ej}\rangle$ remains approximately constant for a dimensionless impact speed larger than ~68, and *Rioul et al.* [2000] and *Beladjine et al.* [2007] reported similar results.

However, Eq. (23) is only valid for large impact speeds, where $N \gg 1$, such that momentum and energy conservation are automatically satisfied by the statistical (ensemble) approach of Eqs. (18, 20). For smaller impact speeds, for which $N \sim 1$, the speed of ejected particles can no longer be approximated by Eq. (23), because momentum and energy conservation do not allow the high-speed tail of the exponential distribution of impact speeds of Eq. (18) with $\langle v_{ej}\rangle$ defined by Eq. (23). Thus, for smaller impact speeds, the discrete nature of the ejection process (that is, $N \approx 1$ rather than $N \gg 1$) provides explicit constraints on momentum and energy conservation that are not automatically satisfied by Eqs. (18) and (20),

$$\sum_i m_{ej}^i v_{ej}^i \leq (1-\alpha_{reb})m_{imp}v_{imp}, \text{ and} \tag{24a}$$

$$\sum_i m_{ej}^i v_{ej}^{i\,2} \leq (1-\alpha_{reb}^2)m_{imp}v_{imp}^2, \tag{24b}$$

where the superscript *i* sums over all the ejected particles, and where we again used that $\varepsilon_{reb} = \alpha_{reb}^2$. When the impacting particle has only enough energy to at most eject one surface particle, Eq. (24) thus truncates the probability distribution of ejection speeds given by Eq. (18). This leads to a decrease in the average ejected particle speed for small impact speeds, as was indeed found by numerical [*Anderson and Haff*, 1988, 1991] and experimental studies with natural sand [*Willets and Rice*, 1985, 1986, 1989; *Rice et al.*, 1995]. Note that the constraints of energy and momentum conservation described by Eq. (24) are automatically satisfied in Eqs. (18) and (20) when $N \gg 1$.

Figure 5 compares $\langle v_{ej}\rangle$ obtained from a Monte Carlo simulation using Eqs. (18, 21, 22, 24) with results from experimental [*Willetts and Rice*, 1985, 1986, 1989; Rice et al., 1995] and numerical [*Anderson and Haff*, 1988, 1991] studies. The increase of $\langle v_{ej}\rangle$ at low $v_{imp}$ is reproduced by our analytical model, as is the independence of $\langle v_{ej}\rangle$ for larger $v_{imp}$ reported in the literature [*Werner*, 1987, 1990; *Haff and Anderson*, 1993; *Rioual et al.* 2000; *Oger et al.*, 2005; *Beladjine et al.*, 2007]. The average dimensionless ejection speed presented in Figure 5 can be described by the expression

$$\frac{\langle v_{ej}\rangle}{\sqrt{gD}} = \frac{\langle \alpha_{ej}\rangle}{a}\left[1 - \exp\left(-\frac{v_{imp}}{40\sqrt{gD}}\right)\right], \tag{25}$$

such that Eq. (23) is retrieved for very large dimensionless impact speeds, where $N \gg 1$.

Eq. (25) thus constitutes a physically-based expression of the speed of ejected particles, which shows good agreement with experiments (Figure 5). The distribution of ejection speeds for the whole range of $N$ is well-described by the exponential distribution of Eq. 18, with $\langle v_{ej}\rangle$ given by Eq. 25.

### 2.2.3 Ejection angle of splashed surface particles

Since the collision of soil particles with the surface converts horizontal momentum into vertical momentum, there are no convenient energetic constraints on the angles at which particles are ejected. We therefore use the consensus result of laboratory and numerical studies that the angle at which particles are ejected can be described by an exponential distribution with a mean of 50 degrees from horizontal [*Willetts and Rice*, 1985, 1986, 1989; *Anderson and Haff*, 1988, 1991; *Werner*, 1990; *McEwan and Willetts*, 1991; *Rice et al.*, 1995, 1996].

### 2.2.4 Ejection of particles from mixed soils

The above analysis for the splash function can be easily extended to mixed soils by assuming that a particle's chance of being ejected from the surface depends on its cross-sectional area [*Rice et al.*, 1995; *Shao and Mikami*, 2005]. For a mixed soil, the number of particles ejected from each particle size bin then becomes

$$N^k = \frac{a}{\sqrt{gD}}\frac{m_{imp}}{m_{ej}^j}\left(\frac{D_{ej}^k}{D_{imp}}\right)^2 v_{imp} f^k = \frac{a}{\sqrt{gD}}\frac{D_{imp}}{D_{ej}^k} v_{imp} f^k, \tag{26}$$

where $D_{imp}$ and $D_{ej}^k$ are the diameter of the impacting and ejected particles, and $f^k$ denotes the mass fraction of the $k^{th}$ particle bin of the soil's particle size distribution.

## 2.3 Wind profile

The wind profile over an aerodynamically rough surface in the absence of saltating particles [*Prandtl*, 1935; *Bagnold*, 1941] is given by

$$\overline{U_x}(z) = \frac{u^*}{\kappa}\ln\left(\frac{z}{z_0}\right), \tag{27}$$

where $z$ is the vertical distance from the surface, $u^*$ is the wind shear velocity or friction velocity and is a measure of the gradient of the fluid flow field, and $z_0 \approx D/30$ is the surface roughness [*Nikuradse*, 1933], where $D$ is the characteristic size of soil particles.

The initial wind profile given by (27) is modified by the transfer of momentum between the wind flow and saltating particles. The amount of horizontal fluid momentum that fluxes into the saltation layer depends directly on the shearing of the flow, and is equal to the fluid shear stress $\tau = \rho_a u^{*2}$ above the saltation layer. At steady state, this flux of horizontal momentum into the saltation layer is partitioned between saltating particles ($\tau_p$) and the fluid ($\tau_a$), such that [*Raupach*, 1991]

$$\tau = \tau_a(z) + \tau_p(z). \tag{28}$$

The fluid momentum flux $\tau_a(z)$ in the saltation layer is a function of the velocity gradient,

$$\tau_a(z) = \rho_a \left[ \kappa z \frac{\partial \overline{U_x}(z)}{\partial z} \right]^2, \tag{29}$$

and $\tau_a(z) = \tau$ for $z$ above the saltation layer. Combining Eqs. (28) and (29) then yields

$$\frac{\partial \overline{U_x}(z)}{\partial z} = \frac{1}{\kappa z} \sqrt{u^{*2} - \tau_p(z)/\rho_a}, \tag{30}$$

with the particle momentum flux given by [*Shao*, 2000]

$$\tau_p(z) = \sum_i m^i v_x^i(z) - \sum_j m^j v_x^j(z), \tag{31}$$

where the superscripts $i$ and $j$ respectively sum over all descending and ascending particles that pass the height $z$ per unit area and unit time.

We calculate $\tau_p(z)$ as a function of the particle trajectories (see Section 2.1) and the concentration of saltating particles (see below), and use it to numerically integrate Eq. (30) to obtain the wind profile in the saltation layer

## 2.4 Particle concentration

The concentration of saltating particles is affected by both the capture of impacting saltating particles by the soil bed (Eq. 15) and the production of new saltating particles through splashing (Eq. 26). The concentration $n^k$ of saltating particles in the particle bin $k$ is thus described by

$$\frac{dn^k}{dt} = \sum_i \frac{a}{\sqrt{gD}} \frac{D_{\text{imp}}^i}{D_{\text{ej}}^k} v_{\text{imp}}^i f^k - \sum_{j_k} 1 - B \left[ 1 - \exp\left( -\wp v_{\text{imp}}^{j_k} \right) \right], \tag{32}$$

where $i$ and $j_k$ respectively sum over all saltating particles and over all particles in bin $k$ that are impacting the soil surface per unit time and unit area. The first term on the right-hand side accounts for the production of saltating particles through splashing, and the second term accounts for the loss of saltating particles to the soil. As the model progresses through successive iterations (see Figure 2), it uses Eq. (32) to converge to the steady-state particle concentration. Indeed, if the number of splashed surface particles is greater than the number of saltating particles settling back to the soil surface, then the concentration of saltating particles increases. This augments the particle momentum flux and thus decreases the wind speed (Eq. 30), which lowers the typical impact speed of saltating particles, thus reducing the number of splashed particles. If, on the other hand, the number of splashed particles is insufficient to balance the settling of saltating particles back to the soil surface, then the particle concentration will decrease. This increases the wind speed and thus the typical impact speed, which in turn increases the number of splashed particles. The model thus iteratively adjusts the particle concentration until steady-state is reached and the particle concentration remains constant with time (i.e., $dn^k/dt = 0$, for all $k$). In steady-state, we then have that

$$\sum_i \frac{a}{\sqrt{gD}} \frac{D^i_{imp}}{D^k_{ej}} v^i_{imp} f^k = \sum_{j_k} 1 - B\left[1 - \exp\left(-\gamma v^{j_k}_{imp}\right)\right], \quad (33)$$

for all $k$. As mentioned in Section 1, the stochastic nature of the interaction of saltating particles with the soil surface and with the turbulent wind field means that the model reaches a dynamic balance in which Eq. (33) is satisfied over longer time scales (a few seconds; *Anderson and Haff*, 1988, 1991; *Jackson and McCloskey*, 1997). We believe this is an accurate representation of natural saltation.

Since the parameters $a$, $B$, and $\gamma$ in Eqs. (32, 33) have not been accurately determined by measurements (Table 1), a useful constraint on their values is that Eq. (33) must be satisfied at the impact threshold. Since the particle concentration (and thus $\tau_p(z)$ in Eq. (30)) is small at the impact threshold, the wind profile is simply given by Eq. (27), such that particle trajectories are obtained in a straightforward manner. Indeed, for given values of the parameters $a$, $B$, and $\gamma$, we can calculate the value of the impact threshold at which Eq. (33) is satisfied. We find that the functional form of the impact threshold is reproduced almost independently of the values of these parameters, and that $a = 0.020$, $B = 0.96$, and $\gamma = 1.0$ s/m provides good quantitative agreement with measurements of the impact threshold (see Figure 6). These parameter values are in agreement with available laboratory and numerical experiments (Table 1). To our knowledge, no previous numerical saltation model has been able to reproduce measurements of the impact threshold.

An additional constraint on the values of $a$, $B$, and $\gamma$ can be obtained by using Eq. (33) to determine an approximate average impact speed in steady-state saltation. This can be done by assuming that particle impact speeds are exponentially distributed (see Eq. 18), as previous studies have suggested [*Anderson and Hallet*, 1986] and results from our model indicate (not shown). Solving Eq. (33) for the average impact speed in this manner yields $\overline{v_{imp}} \approx 1.2$ m/s for 250 μm particles. Note that assuming different plausible impact speed distributions, such as a gamma function [*White and Schulz*, 1977], yields only slightly different values of $\overline{v_{\mathrm{imp}}}$. Since the average impact speed is independent of shear velocity [*Ungar and Haff*, 1987], we also expect particle speeds for different shear velocities to converge near the surface. Recent measurements of particle speeds using laser-Doppler anemometry in a wind-tunnel [*Rasmussen and Sorensen*, 2008] have indeed found that particle speeds for different shear velocities converge to a common value of 1.3-1.5 m/s at 4 mm from the surface. This agreement between measurements and the qualitative and quantitative predictions of our model supports the physical basis underlying our splash parameterization and the chosen values for the parameters *a*, *B*, and γ.

**2.5 Treatment of particles in creep and suspension**

As mentioned in the introduction, sediment can be transported by wind in suspension (< 70 μm), saltation (~70 – 500 μm), or creep (> 500 μm) [*Shao*, 2000]. Our model implicitly accounts for creep through the splash parameterization (Section 2.2). Indeed, Eq. (24) limits the speed with which a massive particle can be ejected from the surface, which in essence describes the process of impacts of smaller particles 'pushing' a larger surface particle in the direction of the wind flow. The good agreement of our model with measurements of the saltation mass flux profile close to the surface (see section 3.1) thus supports the physical basis of our splash parameterization.

We plan to include the emission and transport of suspended dust in a future version of the model.

**2.6 Discussion of model assumptions**

The wind-driven motion of sand particles over a mobile particle bed is a complex process. As also done in previous studies [e.g.,*Werner*, 1990; *Anderson and Haff*, 1991; *Shao and Li*, 1999], we focus on the most important physical processes and make several assumptions to keep our numerical model of saltation manageable. Below we list and discuss the most important assumptions made in our model.

1. *Steady-state saltation*. When saltation is initiated, the drag of saltating particles on the wind increases the apparent surface roughness [*Owen*, 1964; *McEwan and Willetts*, 1993]. The time scale associated with the adjustment of the near-surface wind to this additional roughness is short – on the order of one second [*Anderson and Haff*, 1988, 1991; *McEwan and Willetts*, 1993; *Jackson and McCloskey*, 1997]. However, the time scale required for the near-surface boundary layer to fully adjust to the flow above the saltation layer is much larger [*McEwan and Willetts*, 1993]. We assume that the flow in the saltation layer is fully adjusted to the flow above the saltation layer, which is not always the case in natural saltation.
2. *Wind speed perpendicular to gravity.* While we define the surface as perpendicular to gravity in the results presented in this article, the model is capable of simulating saltation on sloping terrain. We also assume that the wind flow is parallel to the surface. However, a non-zero flow velocity perpendicular to the surface can be included in the model, as done in a previous numerical model by *Yue and Zheng* [2007].

3. *The soil surface is flat.* Sand ripples with typical heights of ~1 cm [*Bagnold*, 1941] usually form during saltation on dunes and beaches. Such ripples will affect the wind flow. However, we follow previous investigators [e.g., *Anderson and Haff*, 1988, 1991; *Shao and Li*, 1999; *Almeida et al.*, 2006] and for simplicity assume that the soil surface is flat.
4. *Particle motion is modeled in two dimensions only.* We assume particle speed to be zero in the direction perpendicular to the plane spanned by the wind and gravitational vectors, while experiments show that ejected and rebounding particles have a small but non-zero speed in this direction [*Xie et al.*, 2007]. Neglecting this component of the particle momentum slightly affects the splash parameterization of Section 2.2 [*Zheng et al.*, 2008].
5. *Mid-air collisions and electrostatic forces are neglected.* For large shear velocities (i.e., $u^* > $ ~0.5 m/s), the particle concentration becomes so large that saltating particles are likely to collide with one or several other particles during a single hop [*Sorensen and McEwan*, 1996; *Dong et al.*, 2005; *Huang et al.*, 2007]. Moreover, electric forces due to sand electrification become large enough to affect particle lifting [*Kok and Renno*, 2006] and trajectories [*Kok and Renno*, 2008]. Since both these processes are of much less importance for small to medium wind shear velocities (i.e., $u^* < $ ~0.5 m/s), we do not include these processes in the version of the model introduced in this article. However, work is in progress to include sand electrification and mid-air collisions in a future model version [*Kok and Renno*, 2009].

## 3 Testing of the model with measurements

We test our model by comparing its results to measurements of the horizontal and vertical profiles of particle mass flux, the total height-integrated mass flux, the size distribution of saltating particles, and the wind profile and aerodynamic roughness length during saltation. When available, we use field measurements rather than wind tunnel measurements since recent studies have shown wind tunnel measurements to differ significantly and systematically from natural saltation [*Farrell and Sherman*, 2006; *Sherman and Farrell*, 2008].

The values of the parameters used in the model are listed in Table 1. We have also included a subjective estimate of the uncertainty of these parameters, as well as a relative indication of the model sensitivity. We hope these estimates can help guide future experimental studies of saltation.

### 3.1 Particle mass flux profiles

Detailed field measurements of the variation of the particle mass flux with height were made by several investigators and are summarized in *Farrell and Sherman* [2006]. Our model shows good agreement with such vertical mass flux profiles as measured by *Greeley et al.* [1996] and *Namikas* [2003] for low ($u^* = 0.31$ m/s) and medium ($u^* = 0.48$ m/s) shear velocities (Figure 7a, b). For larger shear velocities ($u^* = 0.63$ m/s), our model appears to underestimate the decrease in mass flux with height (Figure 7c). A possible reason for this is the absence in the present model version of electrostatic forces, which are thought to decrease the height of particle trajectories as the wind speed increases [*Kok and Renno*, 2008]. Detailed measurements of the horizontal profile of the particle mass flux (i.e., the variation of the particle deposition rate with horizontal distance from a certain starting point) have also been made by *Namikas* [2003]. Simulations with our model show excellent agreement with these measurements (Figure 7d-f).

Figure 8 compares modeled and measured horizontal and vertical mass flux profiles of particles of various sizes [*Namikas*, 2006]. There is reasonable to good agreement between measurements and the predictions of our model, especially when the many uncertainties that affect the results are considered. The predicted flux of fine particles (< ~200 μm) does however decay somewhat too quickly with vertical and horizontal distances (Figure 8a, d). These particles are substantially affected by turbulence [*Anderson*, 1987] and this discrepancy could thus be an indication that the modeled Lagrangian time scale (see Section 2.1.2) is too short. Field measurements of this time scale in the saltation layer would thus be a useful addition to the literature.

Another possible explanation for this discrepancy could be that smaller particles rebound with a greater fraction of their inbound kinetic energy than larger particles do. Indeed, *Namikas* [2006] recently proposed that particles leave the surface with a kinetic energy that is independent of particle size. A simple model using this assumption shows good agreement with measurements [*Namikas*, 2006]. However, this model requires the speed of small particles leaving the surface to be several times their terminal speed, which would imply that these particles actually gain energy upon rebounding from the surface. This is energetically inconsistent. Moreover, results of a wide range of laboratory experiments have consistently reported that the speed with which particles leave the surface is a constant fraction of the impact speed, and that this fraction is independent of particle size [*Willetts and Rice*, 1985, 1986, 1989; *Rice et al.*, 1995, 1996; *Wang et al.*, 2008] and impact speed [*Rioual et al.*, 2000; *Oger et al.*, 2005; *Beladjine et al.*, 2007]. Nonetheless, a more comprehensive investigation of Namikas' hypothesis is desirable.

**3.2 Height-integrated mass flux**

The total height-integrated mass flux of saltating particles is a key parameter for studies of dune formation [*Sauermann et al.*, 2001], wind erosion [*Sterk*, 2003], and dust aerosol emission [*Marticorena and Bergametti*, 1995]. Many wind-tunnel and field measurements have therefore measured the variation of the total mass flux with shear velocity. These measurements are however difficult to compare directly because of variations in experimental conditions, such as particle size, wind-tunnel characteristics, and air pressure. To nonetheless make a comparison between the large body of experimental studies of saltation mass flux and our model predictions, we non-dimensionalize the total mass flux [*Iversen and Rasmussen*, 1999],

$$Q_0 = \frac{\rho_a Q}{g u^{*3}}, \qquad (34)$$

where $Q$ is the total height-integrated saltation mass flux, which is usually assumed to scale with the cube of the shear velocity [*Bagnold*, 1941; *Owen*, 1964; *Iversen and Rasmussen*, 1999].

Figure 9 compares our model predictions to a compilation of field and wind-tunnel measurements of the dimensionless mass flux [Iversen and Rasmussen, 1999]. Our model reproduces the observed peak of the dimensionless mass flux at $u*/u*_{it} \approx 2$ [Iversen and Rasmussen, 1999], where $u*_{it}$ is the impact threshold, as well as the subsequent decrease for larger shear velocities. Many empirical models are unable to reproduce these features (see Figure 9 and *Iversen and Rasmussen*, 1999). The predicted height-integrated mass flux does appear larger than reported by most experimental studies, which may be because sand collectors used in these studies have an efficiency of only ~50-70 % [*Greeley et al.*, 1996; *Rasmussen and Mikkelsen*, 1998]. Moreover, both mid-air collisions and strong electrostatic forces are hypothesized to decrease the mass flux at large shear velocities [*Sorensen and McEwan*, 1996; *Sorensen*, 2004; *Kok and Renno*, 2008]. Since both these processes are not included in the present model version, the overestimation of the mass flux at large shear velocities is thus expected.

## 3.3 Size distribution of saltating particles

Once saltation is initiated, the transfer of momentum to the soil bed by particle impacts causes a wide range of particle sizes to enter saltation. Thus, saltation is not limited to those particles whose threshold shear velocity ($u*_t$) is below the wind shear velocity ($u*$), as is often assumed [e.g., *Marticorena and Bergametti*, 1995]. Rather, the size distribution of saltating particles is determined by two factors: (i) the probability of particles of a given size to be ejected from the surface (see Eq. 26), and (ii) the time that particles of a given size spend in saltation before settling back onto the soil surface.

Measurements of the size distribution of saltating particles were reported by *Williams* [1964]. Moreover, we used the size-resolved vertical mass flux profiles reported by *Namikas* [2006] to obtain the saltation size distribution in his field measurements [*Namikas*, 1999, 2003]. The model-predicted saltation size distribution shows good agreement with the measurements of *Williams* [1964] and with those reconstructed from *Namikas* [2006] (Figure 10). In general, we find that the size distribution of saltating particles in the range 100 – 500 μm roughly matches the parent soil size distribution [*Kok and Renno*, 2008]. This occurs because while larger particles have an increased chance of being ejected from the surface (see Eq. 26 and [*Rice et al.*, 1995]), they also tend to have shorter lifetimes. Conversely, smaller particles are ejected less frequently, but have longer lifetimes once ejected. These two effects cause the saltation size distribution to be similar to that of the soil in the range 100 – 500 μm.

Note that both measurements and our model predictions show that the size distribution shifts slightly towards larger particles as the shear velocity increases. The likely physical reason for this phenomenon is that, while the average impact speed stays approximately constant with increasing shear velocity (see discussion in Section 2.4), we find that the probability distribution of impact speeds broadens with shear velocity. As a result, an increasing fraction of impacting particles have very large impact speeds. Since larger surface particles require greater impact speeds to be splashed into saltation, rather than creep along the surface, the number of large particles entering saltation increases with shear velocity. This leads to the observed and predicted slight shift in the saltation size distribution towards larger particle sizes as the shear velocity increases.

## 3.4 The wind speed and roughness length in saltation

Measurements of the wind speed in saltation were made by numerous researchers and are summarized in *Sherman and Farrell* [2008]. Figure 11 shows wind speeds predicted by our model and compared to wind speeds measured on a desert dune by *Bagnold* [1938] and on a beach by *Namikas* [1999]. The model is in reasonable agreement in both cases, but underestimates the wind speed in comparison with *Bagnold* [1938], while it overestimates the wind speed in comparison with *Namikas* [1999]. Note that the focusing of the wind profiles (the so-called 'Bagnold focus' [*Bagnold*, 1936]) at a height of ~1 cm is reproduced in both cases.

At a given shear velocity, the wind speed directly above the saltation layer is determined by the increase in the aerodynamic roughness length produced by the transfer of wind momentum to saltating particles [*Owen*, 1964]. Several models have been proposed to relate the aerodynamic roughness length in saltation to the shear velocity [*Charnock*, 1955; *Raupach*, 1991; *Sherman*, 1992]. However, the most physically plausible relationship is probably the modified Charnock relationship [*Sherman*, 1992; *Sherman and Farrell*, 2008]

$$z_{0S} = z_0 + C_m \frac{(u^* - u^*_{it})^2}{g}, \tag{35}$$

where $z_{0S}$ is the aerodynamic roughness length during saltation, and $u^*_{it}$ is the impact threshold. *Sherman and Farrell* [2008] used a compilation of 137 wind profiles from field measurements and determined the value of the modified Charnock constant to be $C_m = 0.132 \pm 0.080$. However, for a compilation of 197 wind tunnel experiments, they found that $C_m = 0.0120 \pm 0.0007$. This significant difference in the saltation roughness length between field and wind tunnel experiments indicates that most wind tunnel experiments do not successfully replicate the physics of natural saltation [*Sherman and Farrell*, 2008]. A similar result was obtained by *Farrell and Sherman* [2006], who reported that vertical mass flux profiles in wind tunnel experiments are significantly different from those occurring in natural saltation.

Figure 12 compares the model-predicted saltation roughness length with a collection of field measurements compiled by *Sherman and Farrell* [2008]. Our model reproduces the functional form of the modified Charnock model [*Sherman*, 1992] very well, while the agreement with alternative models, such as the Raupach model [*Raupach*, 1991] and the normal Charnock model [*Charnock*, 1955], is not as good (not shown). Moreover, the best-fit value of the modified Charnock constant from our model results is $C_m = 0.125$, which is very close to the value obtained by *Sherman and Farrell* [2008]. Our results are thus in excellent agreement with field measurements of the roughness length in saltation and provide strong support for the physical correctness of the modified Charnock relationship [*Sherman*, 1992; *Sherman and Farrell*, 2008].

## 4 Conclusions

We have developed a comprehensive numerical model that can simulate steady-state saltation over mixed soils. Our model explicitly simulates particle trajectories due to gravitational and fluid forces and accounts for the effects of turbulence using a parameterization that we show to produce good agreement with measurements (Figure 3). We derived a physically-based parameterization of the 'splashing' of surface particles by impacting saltating particles that shows good agreement with available measurements (Figure 5), correctly predicts the average impact speed of particles in steady-state saltation (Section 2.4) and, when implemented in our numerical saltation model, reproduces measurements of the impact threshold (Figure 6).

Our numerical model is the first physically-based model that can reproduce a wide variety of experimental data, including vertical and horizontal profiles of particle mass flux (Figures 7 and 8), the total height-integrated mass flux (Figure 9), the size distribution of saltating particles (Figure 10), and the wind speed in saltation (Figure 11). Our model is also the first to reproduce measurements of the aerodynamic roughness length in saltation (Figure 12) and reproduces the most physically plausible functional form of the dependence of the roughness length on the shear velocity [*Sherman and Farrell*, 2008].

At large shear velocities, there seems to be less agreement between model predictions and measurements of the vertical profile of the mass flux and the total mass flux (Figures 7c and 9). This probably occurs because the current model version neglects mid-air collisions and electrostatic forces, which are both thought to become important at large shear velocities [*McEwan and Sorensen*, 1996; *Kok and Renno*, 2006, 2008]. Work is in progress to include these processes in a future model version [*Kok and Renno,* 2009].

Since we designed our model to use a minimum of empirical relations, we argue that it is a 'general' model that can be applied, with minimal adaptation, to similar problems in different physical regimes, such as saltating snow, saltation on different planets, and saltation in water. Our model is freely available by contacting the first author (J.K.).

As we outlined in the introduction, a detailed physical understanding of saltation is vital to a variety of problems across scientific disciplines. Of particular interest is the emission of dust aerosols by the impacts of saltating particles on the soil surface [*Shao et al*, 1993; *Marticorena and Bergametti*, 1995; *Shao*, 2000]. These dust aerosols substantially affect the Earth's radiative balance through a variety of processes, and understanding the physical mechanism of their emission is thus essential to understanding past and predicting future climate changes [*Sokolik et al.*, 2001; *IPCC*, 2007]. We therefore aim to use our model to develop a physically based parameterization of dust emission for use in climate models.

## 5 Acknowledgements

We thank Steven Namikas and Keld Rasmussen for insightful discussions, Eugene Farrell for providing us with his compilation of aerodynamic roughness lengths, and Shanna Shaked for comments on the manuscript. This research was supported by NSF award ATM 0622539 and by a Rackham Predoctoral Fellowship to J.K.

*Letters*, *100*(1), 014501.

Kok, J. F., and N. O. Renno (2009), The electrification of wind-blown sand on Mars and its implications for atmospheric chemistry, *Geophysical Research Letters*, submitted.

Kok, J. F., and N. O. Renno (2009), *in preparation*.

Lettau, K., and H. H. Lettau (1978), Experimental and micro-meteorological field studies of dune migration. In: Exploring the World's Driest Climate (Ed. by H.H. Lettau and K. Lettau), IES Report, *101*, 110-147. University of Wisconsin-Madison, Institute for Environmental Studies.

Leuning, R., et al. (2000), Source/sink distributions of heat, water vapour, carbon dioxide and methane in a rice canopy estimated using Lagrangian dispersion analysis, *Agricultural and Forest Meteorology*, *104*(3), 233-249.

Loth, E. (2008), Lift of a solid spherical particle subject to vorticity and/or spin, *Aiaa Journal*, *46*(4), 801-809.

Marticorena, B., and G. Bergametti (1995), Modeling the atmospheric dust cycle. 1. Design of a soil-derived dust emission scheme, *Journal of Geophysical Research-Atmospheres*, *100*(D8), 16415-16430.

McEwan, I. K., and B. B. Willetts (1991), Numerical model of the saltation cloud, *Acta Mech. Suppl. 1*, 53-66.

McEwan, I. K., and B. B. Willetts (1993), Adaptation of the near-surface wind to the development of sand transport, *Journal of Fluid Mechanics*, *252*, 99-115.

McLaughlin, J. B. (1991), Inertial migration of a small sphere in linear shear flows, *Journal of Fluid Mechanics*, *224*, 261-274.

Mitha, S., et al. (1986), The grain-bed impact process in aeolian saltation, *Acta Mechanica*, *63*(1-4), 267-278.

Namikas S. L. (1999), Aeolian saltation: field measurements and numerical simulations, PhD Thesis, University of Southern California, Los Angeles.

Namikas, S. L. (2003), Field measurement and numerical modelling of aeolian mass flux distributions on a sandy beach, *Sedimentology*, *50*(2), 303-326.

Namikas, S. L. (2006), A conceptual model of energy partitioning in the collision of saltating grains with an unconsolidated sediment bed, *Journal of Coastal Research*, *22*(5), 1250-1259.

Nemoto, M., and K. Nishimura (2004), Numerical simulation of snow saltation and suspension in a turbulent boundary layer, *Journal of Geophysical Research-Atmospheres*, *109*(D18), D18206.

Nikuradse, J. (1933), Laws of flow in rough pipes (1950 translation), Tech. Memo. 1292, *Natl. Advis. Comm. on Aeronaut.*, Washington, D. C.

Nishimura, K., and J. C. R. Hunt (2000), Saltation and incipient suspension above a flat particle bed below a turbulent boundary layer, *Journal of Fluid Mechanics*, *417*, 77-102.

Oger, L., et al. (2005), Discrete Element Method studies of the collision of one rapid sphere on 2D and 3D packings, *European Physical Journal E*, *17*(4), 467-476.

Owen, P. R. (1964), Saltation of uniform grains in air, *Journal of Fluid Mechanics*, *20*(2), 225-242.

Painter, T. H., et al. (2007), Impact of disturbed desert soils on duration of mountain snow cover, *Geophysical Research Letters*, *34*(12), L12502.

**TABLES**

| Variable (units) | Physical meaning | Relevant literature | Range in literature | Value used in model | Relative uncertainty | Relative sensitivity |
|---|---|---|---|---|---|---|
| $\langle \alpha_{ej} \rangle$ | Average fraction of impacting momentum spent on ejecting surface grains | *Rice et al.* [1995] | 0.14 – 0.20 | $\frac{\langle 1 - \sqrt{\varepsilon_{reb}} \rangle}{2.5} \approx 0.15$ | Medium | Medium |
| β | The ratio of the Lagrangian and Eulerian time scales | *Anfossi et al.* [2006] | 0.3 – 4 | 1 | High | Low |
| $\langle \varepsilon_{reb} \rangle$ | Average fraction of impacting kinetic energy retained by rebounding particle | *Wang et al.* [2008] | 0.43 – 0.46 | 0.45 | Medium | High |
| γ | Parameter that scales the exponential decay with impact speed of a saltating particle's rebound probability | *Anderson and Haff* [1991] | ~2 | 1 | Very high | Low |
| $\theta_{ej}$ | The mean of the exponential distribution that describes the angle from horizontal with which a surface particle is ejected | *Willetts and Rice* [1985, 1986, 1989]; *Anderson and Haff* [1988, 1991]; *McEwan and Willetts* [1991]; *Rice et al.* [1995, 1996] | 40 - 60° | 50° | Low | Low |
| $\theta_{reb}$ | The mean of the exponential distribution that describes the angle from horizontal with which a saltating particle rebounds | *Magnus and Schulz* [1977]; *Willetts and Rice* [1985, 1986, 1989]; *Anderson and Haff* [1988, 1991]; *McEwan and Willetts* [1991]; *Nalpanis et al.* [1993]; *Rice et al.* [1995, 1996]; *Kang et al.*, [2008] | 25 – 50° | 40° | Low | Medium |
| $\rho_a$ (kg/m³) | Air density – calculated using the ideal gas law with P = 101325 Pa, T = 300 K, and a molar mass of 28.9 grams | N/A | N/A | 1.174 | N/A | N/A |
| $\rho_p$ (g/cm³) | Particle density | N/A | N/A | 2.65 | Very low | Low |
| $\sigma_{\varepsilon_{reb}}$ | Standard deviation of the normal distribution that describes the fraction of kinetic energy that is retained upon rebound | *Wang et al.* [2008] | 0.17 – 0.22 | 0.22 | High | Low |
| $\sigma_{\Omega_p}$ (rev/s) | Standard deviation of the normal distribution that describes the particle spin upon leaving the surface of rebounding or ejected grains | *Chepil and Woodruff* [1963]; *White and Schulz* [1977]; *White* [1982]; *Xie et al.* [2007]; *Zou et al.* [2007] | unclear | 500 | Very high | Very low |
| $\Omega_p$ (rev/s) | Mean of the normal distribution that describes the particle spin upon leaving the surface of rebounding and ejected grains | *Chepil and Woodruff* [1963]; *White and Schulz* [1977]; *White* [1982]; *Xie et al.* [2007]; *Zou et al.* [2007] | 100 – 1000 | 400 | High | Medium |
| *a* | Dimensionless constant that scales proportionality between impact speed and number of ejected particles | *McEwan and Willetts* [1991]; *Rice et al.* [1995, 1996] | 0.01 – 0.05 | 0.02 | Medium | High |
| $b_u$ (m/s) | The standard deviation of the turbulent horizontal wind speed | *Shao* [1995]; *Nishimura and Hunt* [2000] | 2.4 – 2.5 | 2.5 | Low | Very low |
| $b_w$ (m/s) | The standard deviation of the turbulent vertical wind speed | *Hunt and Weber* [1979]; *Shao* [1995]; *Nishimura and Hunt* [2000] | 1.2 – 1.5 | 1.4 | Low | Low |
| *B* (s/m) | Probability that a high-speed particle rebounds upon impacting the soil surface | *Mitha et al.* [1986]; *Anderson and Haff* [1991] | ~0.94 – 0.95 | 0.96 | High | Medium |

**Table 1.** Description of parameters used in the numerical model, with the range given in the relevant literature, the value used in the model, a subjective indication of the uncertainty in the parameter's value, and the relative sensitivity of the model results to variations in the parameter's value.

**FIGURES**

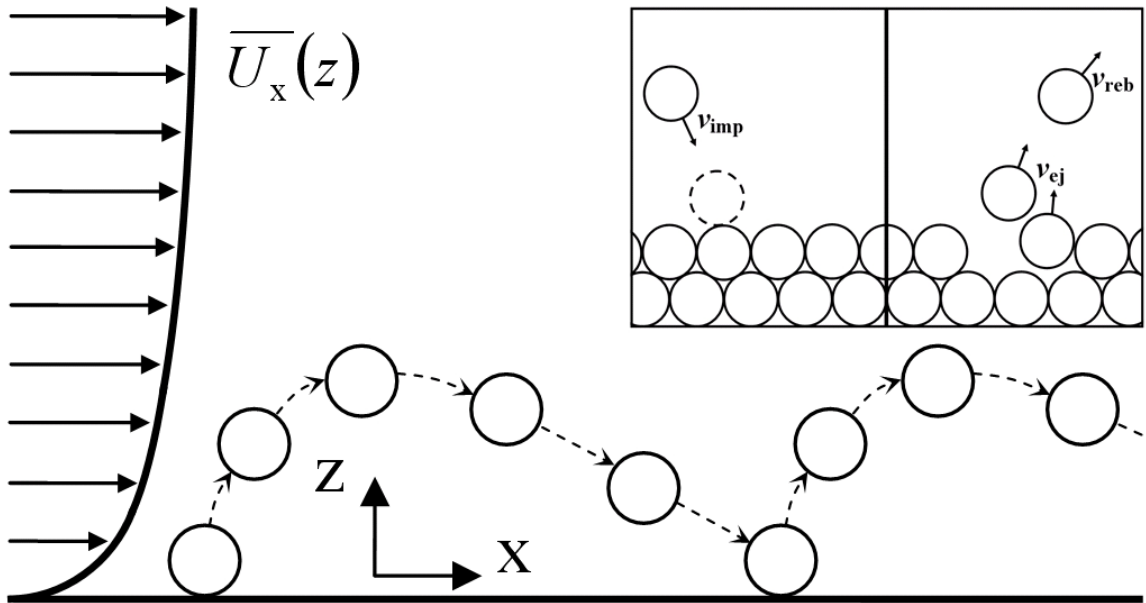

**Figure 1.** Schematic representation of saltation, showing the logarithmic wind profile $\overline{U_x}(z)$ (see Section 2.3) to the left of an idealized spherical sand particle propelled by the wind and bouncing along the surface. After lift-off from the surface, saltating particles gain horizontal momentum from the wind, which is partially converted into vertical momentum after colliding with the surface and rebounding. The inset shows a schematic representation of a saltating particle approaching the soil surface (left) and rebounding from it and ejecting (or 'splashing') several surface particles (right).

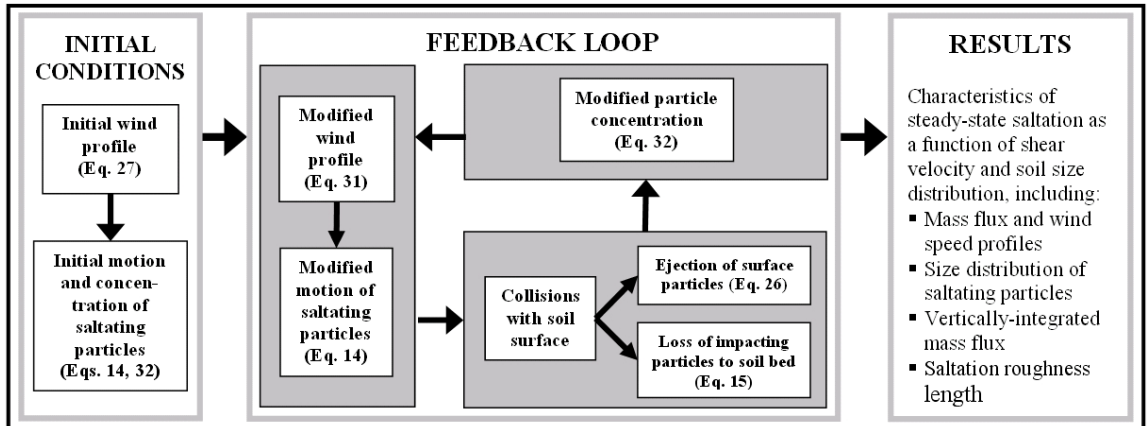

**Figure 2.** Schematic diagram of our numerical saltation model. As in previous studies [*Anderson and Haff*, 1988, 1991; *Werner*, 1990; *McEwan and Willetts*, 1991, 1993], we model saltation by explicitly simulating (i) particle trajectories, (ii) the modification of the wind profile through momentum transfer between the wind flow and saltating particles, and (iii) the collision of particles with the soil surface and the subsequent 'splashing' of surface particles into the fluid stream. The model is initiated by aerodynamically lifting several particles with a speed sufficient to reach a few particle diameters [*Anderson and Haff*, 1991], after which the steps in the feedback loop are repeated until the changes in the saltation trajectories, the wind profile, and the particle concentration are smaller than a specified value in successive iterations. Because of the stochastic interaction of saltating particles with the turbulent wind (Section 2.1.2) and the soil surface (Section 2.2), steady-state saltation as simulated by our model is a dynamic balance over longer timescales. This is also characteristic of natural saltation [e.g., *Anderson and Haff*, 1991; *Jackson and McCloskey*, 1997]. The model does not incorporate aerodynamic lifting in steady-state saltation, because the fluid shear stress at the surface is below the threshold for lifting (see Section 1). For computational efficiency, the model explicitly simulates the trajectories of only a fraction of the particles, and considers those representative of the entire ensemble of saltating particles. Increasing this fraction does not significantly affect the results presented here.

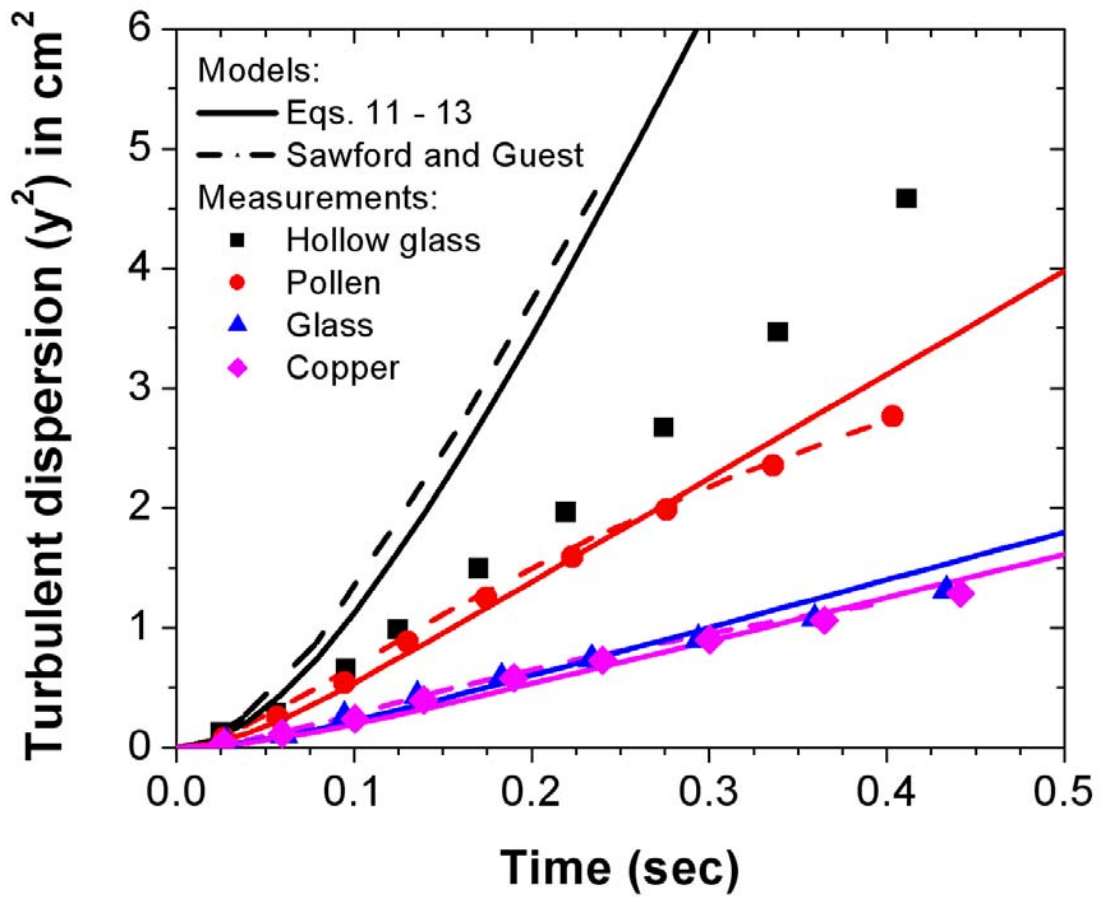

**Figure 3.** Turbulent dispersion perpendicular to the mean flow as measured by *Snyder and Lumley* [1971] for 46.5 μm diameter hollow glass (0.26 g/cm$^3$; black squares), 87.0 μm pollen (1.0 g/cm$^3$; red circles), 87.0 μm solid glass (2.5 g/cm$^3$; blue triangles), and 46.5 μm copper (8.9 g/cm$^3$; magenta diamonds) particles. Included for comparison are the turbulent dispersion simulated for similar particles by the model of *Sawford and Guest* [1991] (dashed black and colored lines) and by Eqs. (11) – (13) (solid black and colored lines). Good agreement between model predictions and measurements can be seen, except for the hollow glass particles which are the lightest of the four kinds of particles and are least characteristic of saltating particles.

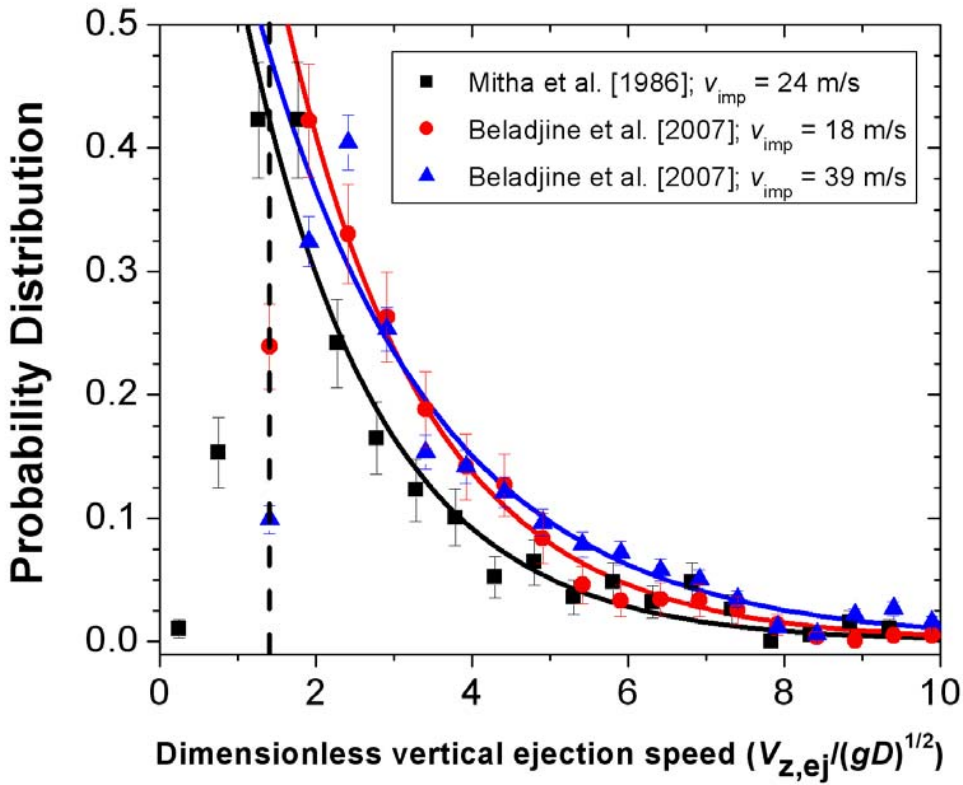

**Figure 4**. Probability distribution of the dimensionless vertical ejection speed. Shown are experimental results for 4 mm steel particles impacting a bed of similar particles at 24 m/s (black squares) [*Mitha et al.*, 1986], and for 6 mm PVC particles impacting at 18 m/s (red circles) and 39 m/s (blue triangles) [*Beladjine et al.*, 2007]. The data above the threshold for which particle detection is reliable (dashed line) [*Beladjine et al.*, 2007] are well-described by exponential distributions (black, red, and blue solid lines). Error bars are derived from the total number of particle counts contained in each data point.

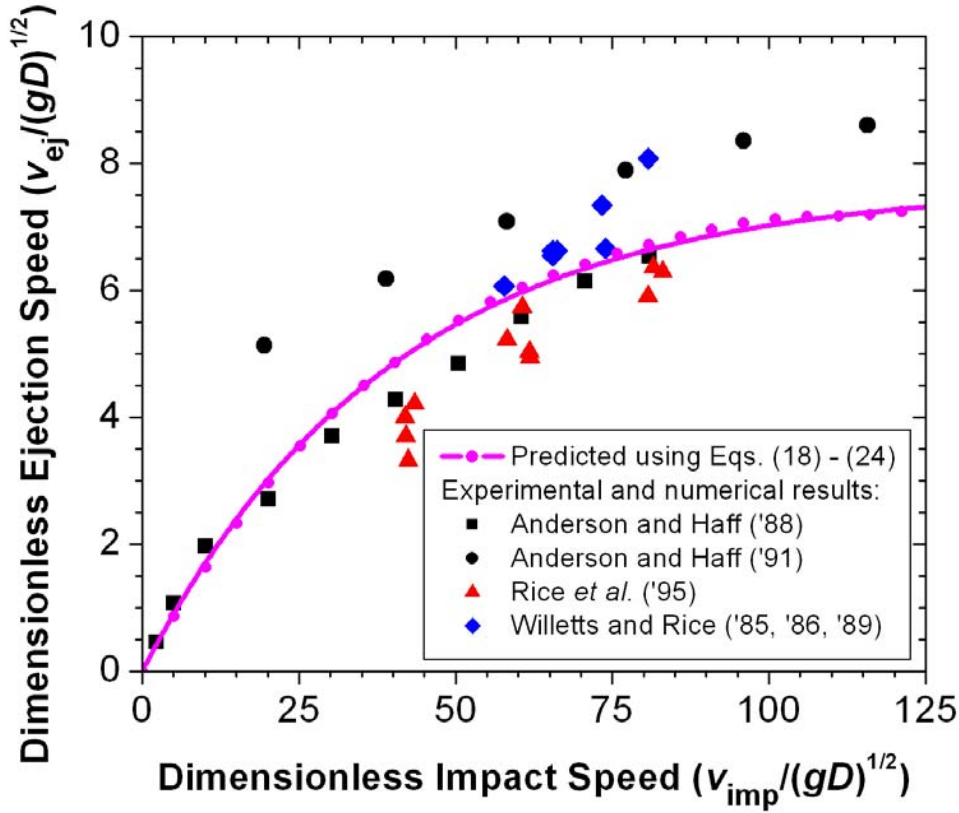

**Figure 5**. The average dimensionless speed of ejected surface particles ($\overline{v_{ej}}/\sqrt{gD}$) as a function of the dimensionless speed of the impacting particle ($v_{imp}/\sqrt{gD}$). We used Eqs. (18) – (24) to perform a Monte Carlo simulation (magenta circles) of particles impacting a bed of similar particles, for which we used parameters as specified in Table 1. The model results do not depend on the particle size. The magenta solid line represents the fit to these results as given by Eq. (25). Experimental results from *Willetts and Rice* [1985, 1986, 1989] (red triangles) denote the average speed of particles splashed from a bed of mixed particles by a medium-sized (250-355 μm) impacting particle, whereas the results from *Rice et al.* [1995] (blue diamonds) represent the average speed by which fine (150-250 μm), medium (250-355 μm), and coarse (355-600 μm) particles are ejected from a bed of mixed particles by an impacting particle of the same size. The numerical studies of *Anderson and Haff* [1988] and [1991] (black squares and circles, respectively) were performed for 2-dimensional sand grains of 1 mm and 230-320 μm diameter, respectively. Results from similar experimental and numerical studies with particles other than sand grains [e.g., *Oger et al.*, 2005; *Beladjine et al.*, 2007] are omitted. The sphericity and the elastic and friction coefficients of such particles differ from those of natural sand, which likely affects the experimental results [*Mitha et al.*, 1986; *Anderson and Haff*, 1991].

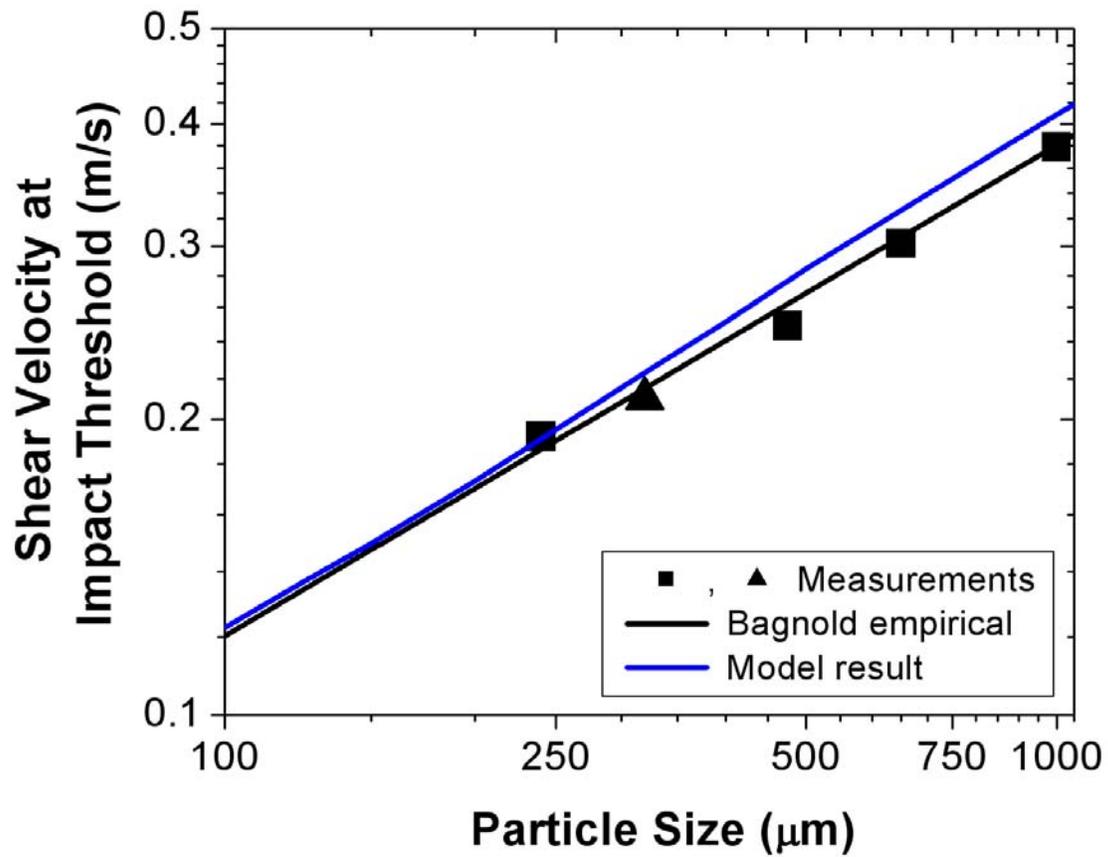

**Figure 6**. Impact threshold for Earth ambient conditions as measured in wind-tunnel experiments by *Bagnold* [1937] (black squares) and *Iversen and Rasmussen* [1994] (black triangle), and predicted by our numerical saltation model (blue line). Also plotted is Bagnold's empirical relation for the impact threshold (black line) [*Bagnold*, 1937, pp. 435].

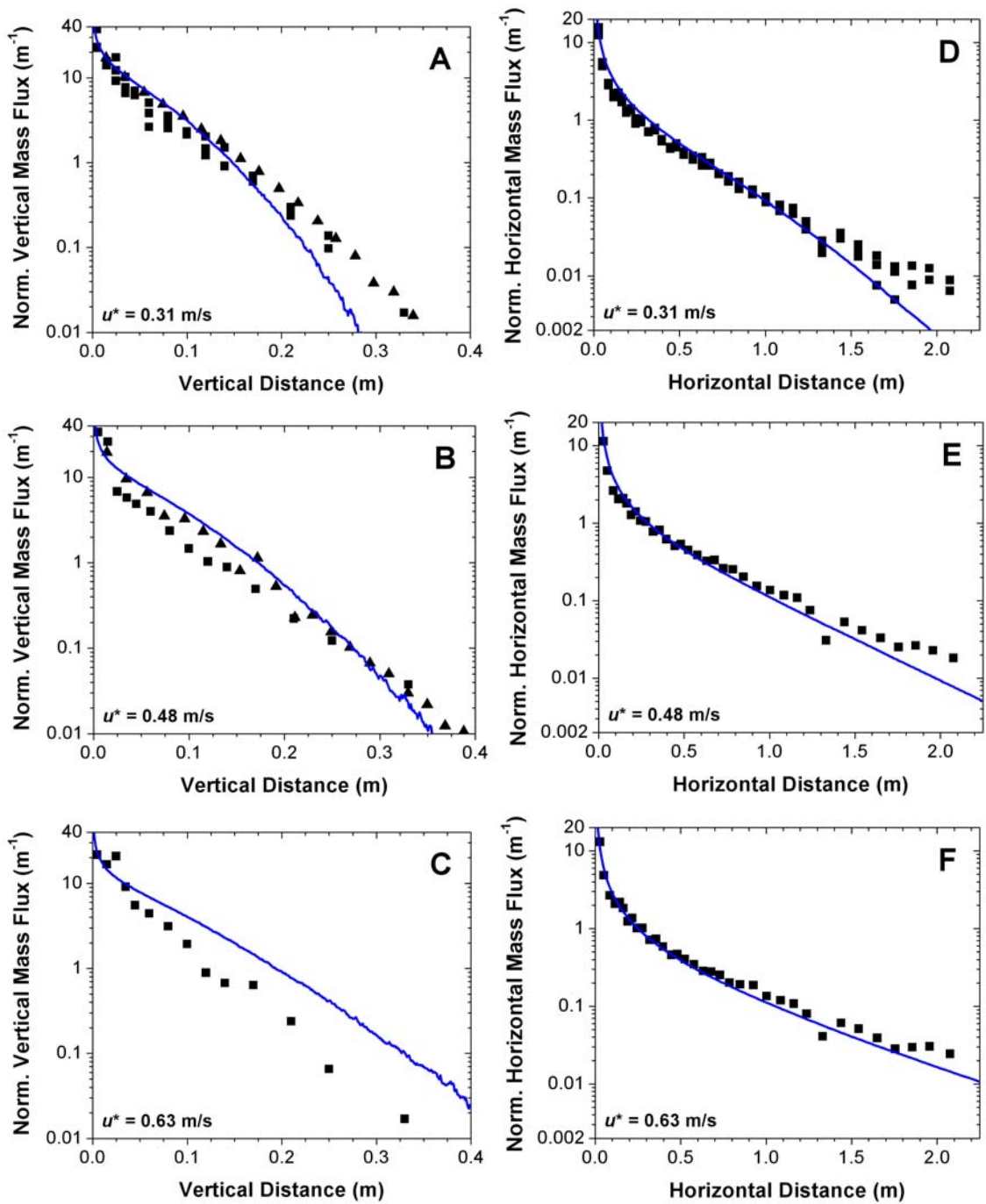

**Figure 7**. Vertical and horizontal mass flux profiles for $u^* = 0.31$, $0.48$, and $0.63$ m/s. Triangles denote vertical mass flux profile measurements from runs 4 and 5b of *Greeley et al.* [1996] and squares denote both vertical and horizontal mass flux profile measurements from runs 4, 5, 8, 13, and 14 of *Namikas* [2003]. Model results (solid blue line) were obtained for the size distribution reported in *Namikas* [2003], which we assume characteristic for *Greeley et al.*'s measurements as well, since their measurements were taken in a similar location. Both measured and modeled mass flux profiles were normalized by their total mass flux to facilitate comparison.

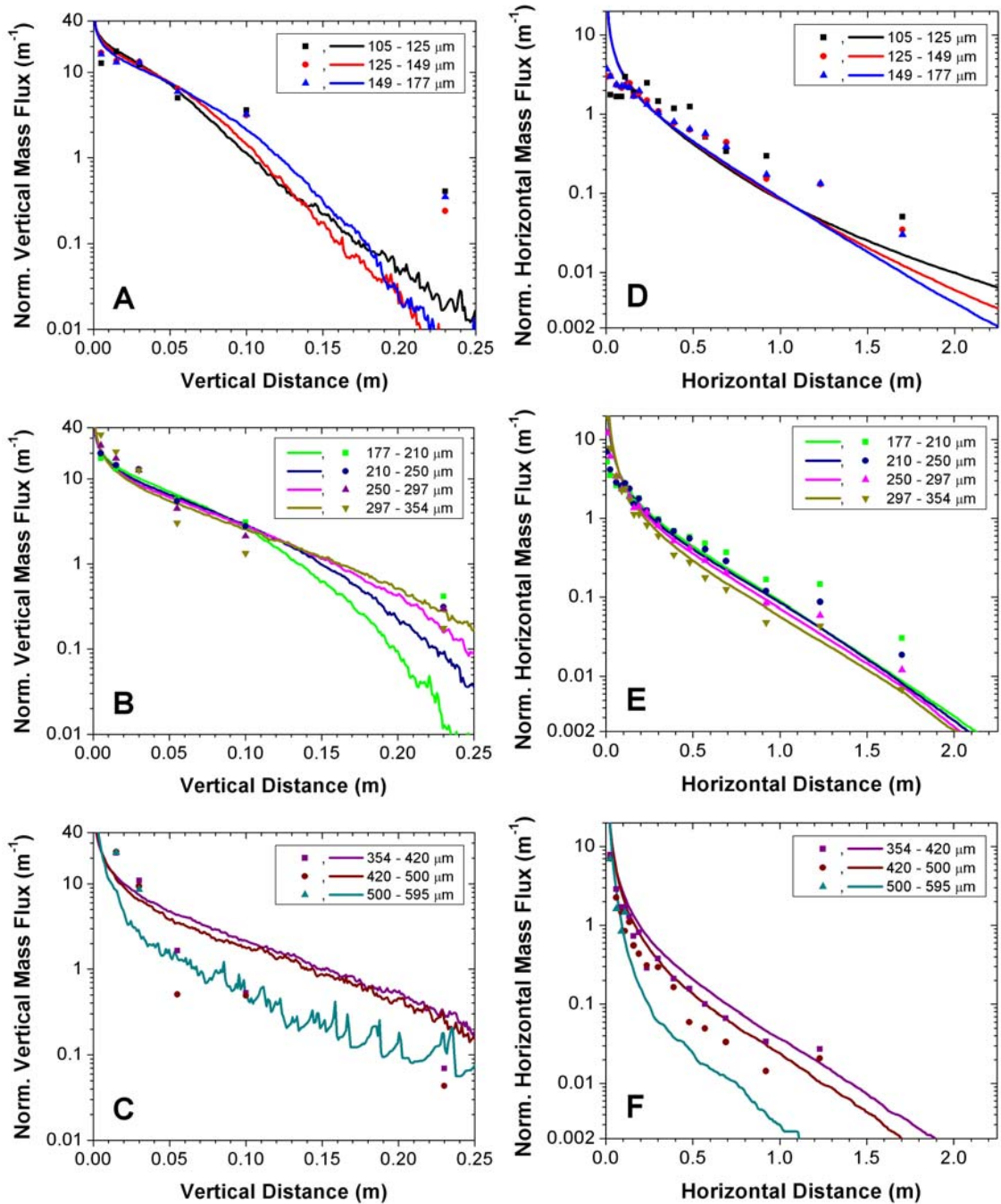

**Figure 8.** Vertical and horizontal mass flux profiles for different particle sizes. The colored symbols represent measurements taken at $u^* = 0.36$ m/s by *Namikas* [1999, 2003, 2006], and colored lines denote the model prediction for the corresponding particle size. In order to facilitate comparison, both measured and modeled mass flux profiles are normalized by the total saltation mass flux of a given particle bin. The increased noise at larger heights in the vertical mass flux profiles is due to the low probability of particles to saltate at those heights, which results in a larger uncertainty.

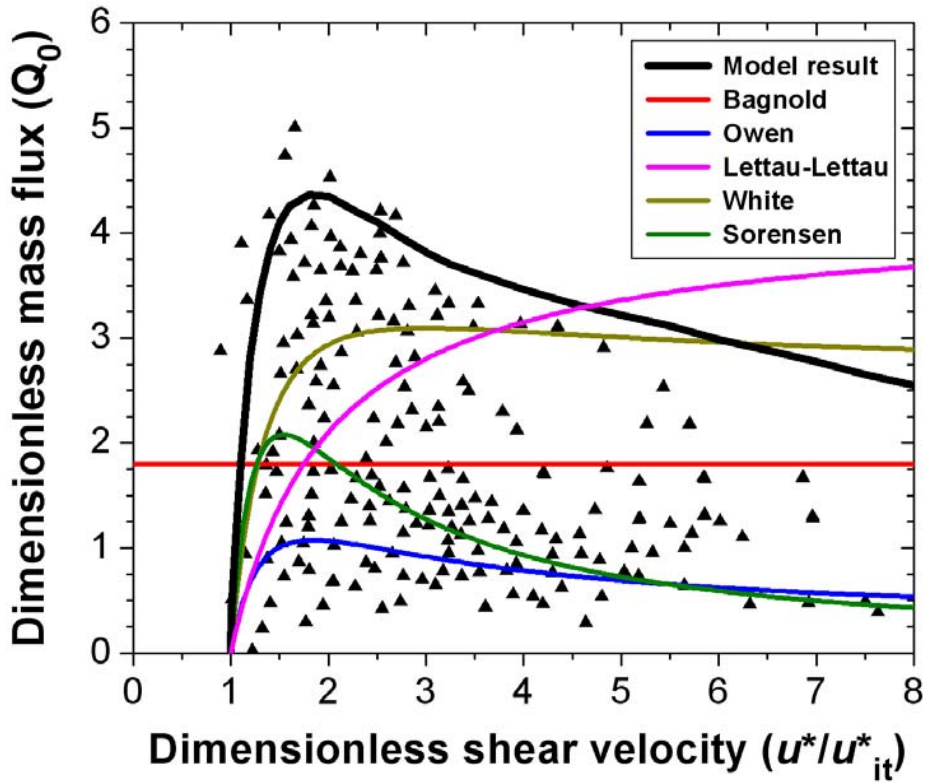

**Figure 9**. Dimensionless saltation mass flux $Q_0$ (see Section 3.2) as a function of dimensionless shear velocity ($u^*/u^*_{it}$, where $u^*_{it}$ is the impact threshold) simulated with our numerical model (black line), and compared with results from over a dozen wind tunnel studies and one field study compiled by *Iversen and Rasmussen* [1999] (triangles). The large scatter in the experimental results is likely caused by varying experimental conditions, such as particle size, air pressure, and wind-tunnel characteristics [*Iversen and Rasmussen*, 1999]. A peak in the dimensionless mass flux is nonetheless apparent around $u^*/u^*_{it} \approx 2$, and is reproduced by the model. For comparison we also included prominent empirical equations of the saltation mass flux (colored lines) by *Bagnold* [1941] ($Q_0 = 1.8$), *Owen* [1964] ($Q_0 = [0.25 + v_t/3u^*][1 - (u^*_{it}/u^*)^2]$), where $v_t$ is the terminal velocity of saltating particles), *Lettau and Lettau* [1978] ($Q_0 = 4.2[1 - u^*_{it}/u^*]$), *White* [1979] ($Q_0 = 2.61[1 - u^*_{it}/u^*][1 + u^*_{it}/u^*]^2$), and *Sorensen* [1991, 2004] ($Q_0 = [1 - u^{*2}_{it}/u^{*2}][\alpha + \gamma u^*_{it}/u^* + \beta u^{*2}_{it}/u^{*2}]$, with $\alpha = 0$, $\beta = 3.9$, and $\gamma = 3.0$ from Figure 3 in *Sorensen* [2004]). Model results (black line) were obtained for the size distribution of typical beach sand reported in *Namikas* [2003], with an approximate median diameter of 250 μm. For very large shear velocities (i.e., $u^*/u^*_{it} > \sim 4$), a substantial fraction (on the order of 5 – 25 %) of the predicted mass flux is due to suspended sand transported at large heights. To exclude this fraction from the saltation mass flux, we omit the mass flux transported above a height of 0.5 meters, in accordance with the vertical extent of mass flux collectors used in wind-tunnel [e.g., *Iversen and Rasmussen*, 1999] and field studies [e.g., *Bagnold*, 1938; *Greeley et al.*, 1996; *Namikas*, 2003].

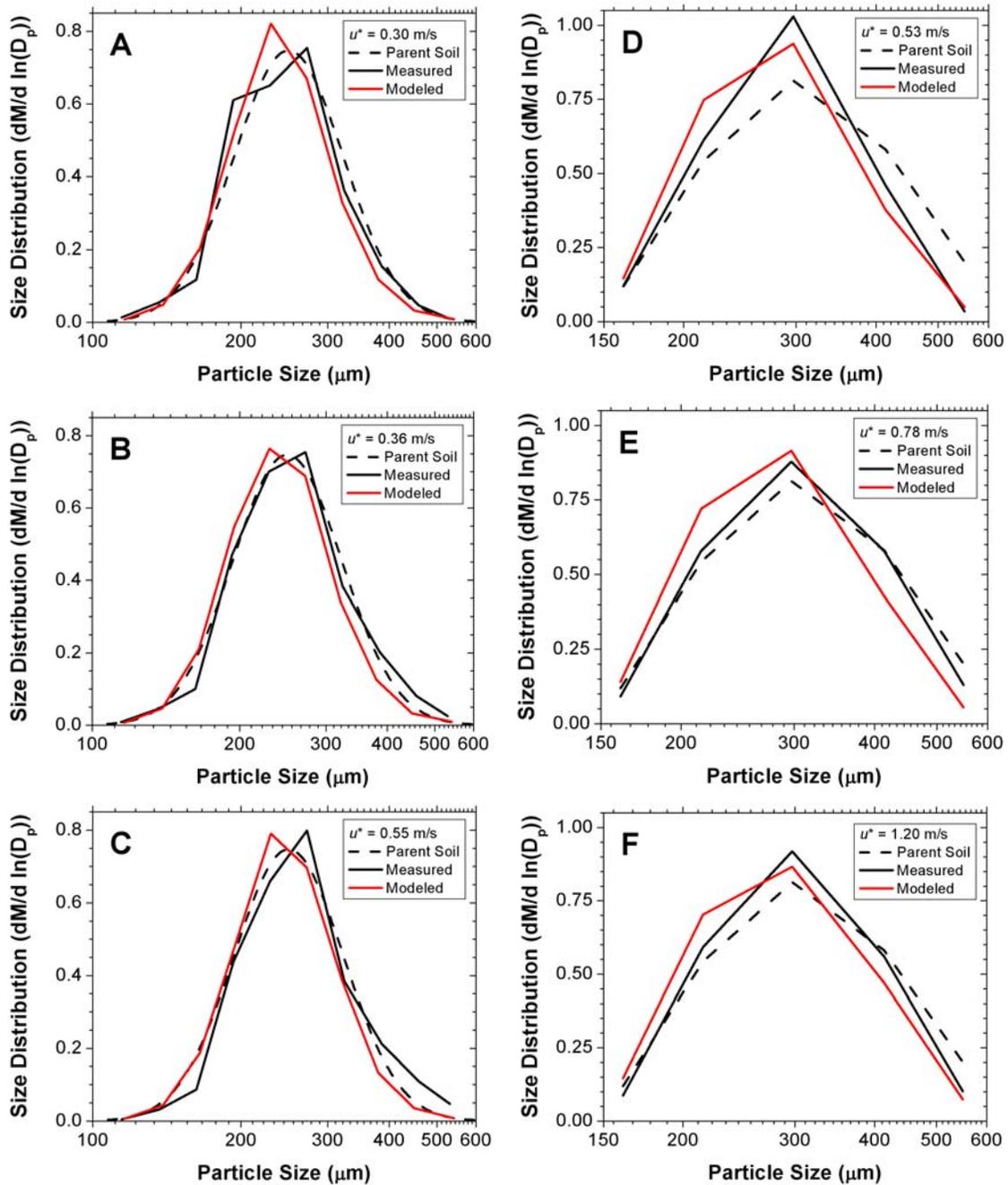

**Figure 10.** Size distributions of saltating particles during saltation, as measured (solid black lines) by *Williams* [1964] (left) and *Namikas* [1999, 2003, 2006] (right) and predicted with our numerical model (red solid lines). Model results were obtained for the same parent soil (dashed black lines) and wind conditions. The saltation size distribution for Namikas' field measurements was obtained by summing the particle size-resolved vertical mass flux reported in Figure 3 of *Namikas* [2006]. We define the size distribution of saltating particles as the contribution of each particle bin to the total height-integrated mass flux, in accordance with measurements [*Williams*, 1964; *Namikas*, 2006].

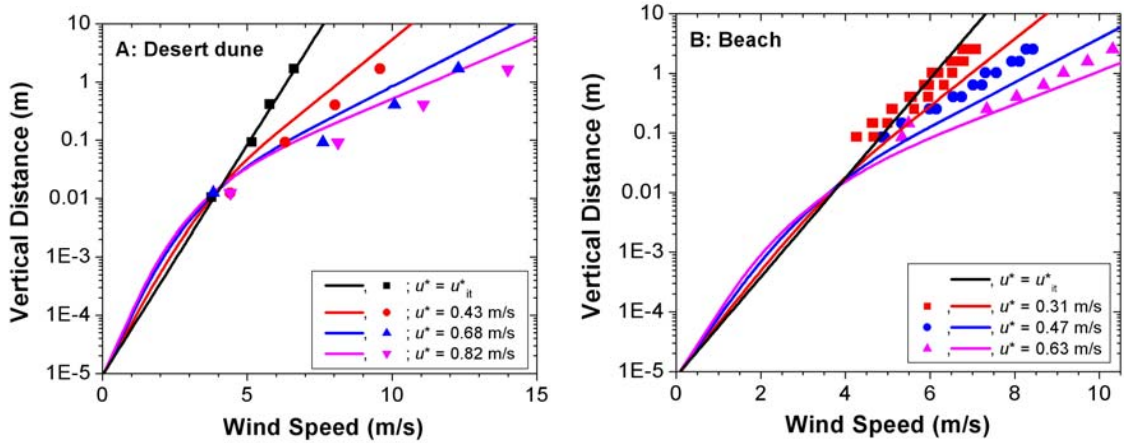

**Figure 11.** Wind profiles during saltation on a desert dune [*Bagnold*, 1938] (symbols in the left graph), on a beach [*Namikas*, 1999] (symbols in the right graph), and modeled (colored lines) for similar conditions. Since *Bagnold* [1938] did not report a soil size distribution, we assume this to be similar to the size distribution of saltating particles (i.e., we used the saltating particle size distribution for $u^* = 0.33$ m/s reported in Bagnold's Figure 7), as experiments indicate (see Figure 9). Using this size distribution, the model predicts an impact threshold (black line) that is in excellent agreement with Bagnold's measured impact threshold (black squares). The model results for *Namikas* [1999] use the size distribution as reported in *Namikas* [2003], for which the model predicts an impact threshold of 0.21 m/s (black line), in good agreement with Namikas' estimated impact threshold of 0.20 – 0.23 m/s [*Namikas*, 1999].

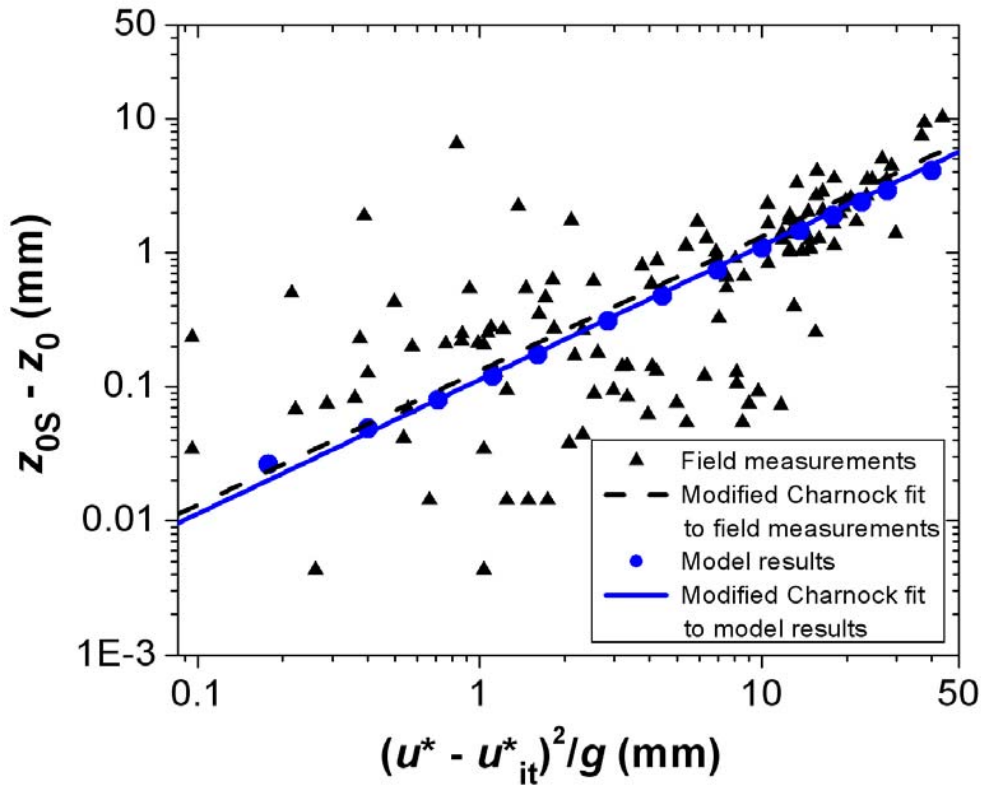

**Figure 12.** Aerodynamic roughness length in saltation from a compilation of field measurements by *Sherman and Farrell* [2008] (black triangles), and simulated by our model (blue circles). Also included are fits with the modified Charnock relationship (Eq. 35) [Charnock, 1955; Sherman, 1992] to the compilation of field measurements (black dashed line) and to our model results (blue solid line). These fit lines nearly overlap (see text) and are therefore difficult to distinguish. The large scatter in the experimental results is probably due to measurement error and variations in experimental conditions, such as particle size, soil moisture content, and surface slope.